\documentclass[a4paper,11pt]{article}
\pdfoutput=1 %

\usepackage{jheppub} %
\usepackage[T1]{fontenc} %
\usepackage{enumitem}

\usepackage{mathrsfs} %
\usepackage{calligra} %
\usepackage{braket} %
\usepackage{slashed} %
\usepackage{cancel}

\usepackage{tikz}
\usetikzlibrary{positioning, calc}
\usetikzlibrary{decorations.markings}

\newcommand{\badat}{\begin{alignedat}}
\newcommand{\eadat}{\end{alignedat}}

\newcommand{\w}{\omega}
\newcommand{\p}{{\partial}}
\newcommand{\zb}{{\bar z}}
\newcommand{\wb}{{\bar w}}

\newcommand{\scri}{{\mathcal I}}

\newcommand{\calD}{{\mathcal D}}

\newcommand{\calI}{{\mathcal I}}
\newcommand{\calR}{{\mathcal R}}

\newcommand{\Y}{{\mathcal Y}}

\title{
The Classical Super-Rotation Infrared Triangle\\
{\small Classical Logarithmic Soft Theorem as Conservation Law in Gravity}}
\author[a]{Sangmin Choi,}\emailAdd{s.choi@uva.nl}
\author[b]{Alok Laddha,}\emailAdd{aladdha@cmi.ac.in}
\author[a]{Andrea Puhm}\emailAdd{a.puhm@uva.nl}

\affiliation[a]{Institute for Theoretical Physics, University of Amsterdam, PO Box 94485, 1090 GL Amsterdam, The Netherlands}
\affiliation[b]{Chennai Mathematical Institute, H1, SIPCOT IT Park, Siruseri, Kelambakkam 603103, India}

\abstract{
The universality of gravitational scattering at low energies and large distances encoded in soft theorems and memory effects can be understood from symmetries. %
In four-dimensional asymptotically flat spacetimes the infinite enhancement of translations, extending the Poincaré group to the BMS group, is the symmetry underlying Weinberg's soft graviton theorem and the gravitational displacement memory effect.  Beyond this leading infrared triangle, loop corrections alter their nature by introducing logarithms in the soft expansion and late time tails to the memory, and this persists in the classical limit.
In this work we give the first complete description of an `infrared triangle' where the long-range nature of gravitational interactions is accounted for.
Building on earlier results \cite{Choi:2024ygx} where we derived a novel conservation law associated to the infinite dimensional enhancement of Lorentz transformations to superrotations, %
we prove here its validity to all orders in the gravitational coupling and show that it implies the classical logarithmic soft graviton theorem of Saha-Sahoo-Sen \cite{Saha:2019tub}. We furthermore extend the formula for the displacement memory and its tail from particles to fields, thus completing the classical superrotation infrared triangle.

}

\begin{document} 
\maketitle
\section{Introduction and summary}

At low energies and large distances gravitational scattering exhibits universal behavior in the form of soft theorems and memory effects that can be traced to an underlying asymptotic symmetry.
The infrared (IR) triangle is an abstraction that encapsulates this universality of infrared physics in classical and quantum scattering. Until recently, complete infrared triangles -- where all three corners and the connection between them is understood -- have only been established for tree-level scattering processes. Loop effects associated to the long-range nature of gravitational interactions crucially modify these infrared relations through novel soft theorems with logarithmic dependence on the energy together with late-time tails in the gravitational field that give rise to so-called tail memory effects. In this work, building on \cite{Choi:2024ygx}, we  complete the first infrared triangle in gravity where these infrared effects must be accounted for: the (classical) superrotation infrared triangle. 

In gravity, the leading infrared triangle encapsulates the realization \cite{He:2014laa} that the gravitational displacement memory formula in classical gravity \cite{Zeldovich:1974gvh, Braginsky:1985vlg, Braginsky:1987kwo, Ludvigsen:1989kg, Christodoulou:1991cr, Wiseman:1991ss, Thorne:1992sdb, Blanchet:1992br, Bieri:2011zb, Tolish:2014bka, Tolish:2014oda} is the classical limit of Weinberg's soft factorisation theorem of gravitational amplitudes \cite{Weinberg:1965nx}. Both these infrared effects are universal in the sense that they do not depend on the details of the interactions. The leading soft graviton theorem and the gravitational displacement memory effect constitute two corners of an infrared triangle. The third corner arises from the realisation \cite{He:2014laa} that both of these infrared effects are nothing but the conservation law (Ward identity) of the supertranslation charge \cite{Bondi:1962px,Sachs:1962wk} of four-dimensional asymptotically flat spacetimes.

Upon reflection an apparent puzzle arises. In four spacetime dimensions the gravitational S-matrix trivialises precisely because of infrared effects which lead to $e^{-\infty}$ suppression factors. How can we then assert that supertranslations are a symmetry of the S-matrix which may not even exist ?! This apparent paradox is resolved in the same spirit in which we interpret Weinberg's soft graviton theorem in four dimensions. Namely, any factorisation theorem is really a statement about the ratio of two regularised amplitudes. Both the numerator and denominator in this ratio are separately ill-defined, but we first regulate them and, in the limit that the regulator is taken to zero, the ratio remains infrared-finite and equals Weinberg's soft factor. 

The subtleties involving infrared effects become more pronounced for the subleading infrared triangle which relates the subleading tree-level soft graviton theorem \cite{Cachazo:2014fwa} to a gravitational spin memory effect \cite{Pasterski:2015tva} both of which can be expressed as a consequence of superrotation symmetry \cite{Barnich:2009se,Barnich:2010ojg}. 
The latter constitutes a local enhancement of Lorentz transformations which act on the celestial sphere at the conformal boundary of asymptotically flat space as local conformal transformations \cite{Kapec:2016jld}.
This has raised the prospect that a putative holographic dual to four-dimensional quantum gravity in asymptotically flat spacetimes may share features with a two-dimensional conformal field theory \cite{He:2014laa,Kapec:2016jld,Nande:2017dba,Donnay:2018neh,Fan:2019emx,Nandan:2019jas,Pate:2019mfs,Adamo:2019ipt,Puhm:2019zbl,Guevara:2019ypd,Kapec:2017gsg,Donnay:2020guq,Kapec:2021eug,Pasterski:2021fjn,Donnay:2022sdg,Pano:2023slc}; over the past few years this has been developed in the celestial holography program \cite{Strominger:2017zoo,Raclariu:2021zjz,Pasterski:2021rjz,McLoughlin:2022ljp,Donnay:2023mrd}.

However, the long-range nature of gravitational interactions casts a veil of ambiguity on the subleading infrared triangle.
Loop-corrections introduce non-analytic terms in the soft expansion in the form of logarithms \cite{Laddha:2018myi}. These render the tree-level subleading soft graviton theorem ambiguous, and thus also its interpretation as the Ward identity associated to superrotation symmetry. Consequently, this will affect %
the conjectured infinite-dimensional $w_{1+\infty}$ symmetry algebra of  quantum gravity in four-dimensional asymptotically flat spacetimes \cite{Guevara:2021abz,Strominger:2021mtt} and  any flat space holographic proposal.

In this work, building on \cite{Choi:2024ygx}, we will show that superrotation symmetry is, in fact, a symmetry of classical gravitational scattering and that its associated Ward identity is the classical {\it logarithmic} soft graviton theorem that was recently established by Saha, Sahoo and Sen \cite{Sahoo:2018lxl,Saha:2019tub}.
For massive point particles it has been shown that the latter can be recast as a {\it tail} to the displacement
memory effect \cite{Laddha:2018vbn,Saha:2019tub,Ghosh:2021bam,Sahoo:2021ctw}; here we  generalize this tail memory result to massive {\it fields} in a formula that is similar to the non-linear memory for massless fields \cite{Favata:2010zu,Laddha:2019yaj}. 
Thus, our work provides the missing symmetry corner of the subleading infrared triangle which explains the universality of the logarithmic soft theorem and the tail memory in the form of superrotation symmetry. We will refer to it as the classical superrotation infrared triangle.

\begin{figure}[ht!]
\begin{center}
  \tikzset{->-/.style={decoration={
    markings,
    mark=at position 0.7 with {\arrow{stealth}},
    mark=at position 0.3 with {\arrowreversed{stealth}}
    },
    postaction={decorate}}
  }
  \begin{tikzpicture}[scale=0.7]
    \coordinate (A) at (0,0);
    \coordinate (B) at (4,0);
    \coordinate (C) at (2,3.464);

    \node[text width=4cm,align=center] at (-2.8,-0.5) {classical logarithmic soft graviton theorem};
    \node[text width=4cm,align=center] at (7.2,-0.5) {tail to the gravitational displacement memory};
    \node[text width=4cm,align=center] at (2,4.1) {superrotation};

    \draw [line width=1.5pt,->-] (A) -- (B);
    \draw [line width=1.5pt,->-] (B) -- (C);
    \draw [line width=1.5pt,->-] (C) -- (A);

  \end{tikzpicture}
  \end{center}
      \caption{Classical superrotation infrared triangle.}
\end{figure}
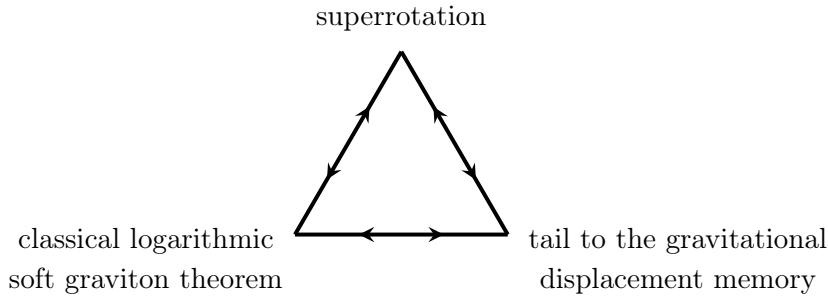

In the remainder of this section we review  salient features of soft theorems in the presence of long-range interactions and present our key findings. In this paper we give detailed proofs of the results first reported in \cite{Choi:2024ygx} including a novel extension of our main result to all orders in the gravitational coupling and we propose a formula for the gravitational memory and its tail for massive fields.

In a companion paper \cite{CLP_sQED} we carry out a similar analysis in scalar QED where we complete the classical superphaserotation infrared triangle associated to the logarithmic soft photon theorem.

\paragraph{Power-law soft theorems.}

Tree-level scattering processes involving a low energy, or {\it soft}, graviton with momentum $k^\mu=\omega q^\mu$ and $N$ particles with {\it hard} momenta $p_i$ with $i=1,...,N$ exhibit tree-level factorization properties expressed as a power-law expansion in the soft graviton energy~$\omega\to 0$ \cite{Weinberg:1965nx,Cachazo:2014fwa,Hamada:2018vrw,Li:2018gnc},
\begin{equation}\label{powersoftexp}
   {\cal M}_{N+1}(p_{1}, \dots, p_{N};(\omega,q,\ell))=\sum_{n=-1}^{\infty} \omega^{n}  {S}_{n}(p_1,\dots, p_N;(q,\ell)){\cal M}_{N}(p_{1}, \dots, p_{N})+\dots\,.
\end{equation} 
The leading \cite{Weinberg:1965nx} and subleading \cite{Cachazo:2014fwa} soft graviton theorems are {\it universal} in that their soft factors only depend on the momenta and angular momenta of the hard particles but not the details of the interactions,
\begin{equation}
  S_{-1}=\frac{\kappa}{2}\sum_{i=1}^N  \frac{\varepsilon_{\mu\nu} p^\mu_i p^\nu_i}{ q\cdot p_i},\quad  S_{0}=-i\frac{\kappa}{2}\sum_{i=1}^N  \frac{\varepsilon_{\mu\nu} p^\mu_i q_\rho J^{\rho\nu}_i}{ q\cdot p_i}.
\end{equation}
While the leading soft theorem is tree-exact, the subleading soft theorem receives corrections at one-loop \cite{Bern:2014oka,Sahoo:2018lxl}. Further subleading soft theorems receive both loop corrections as well as non-universal contributions which, beyond the sub-subleading soft theorem, spoil the factorization property. A more severe issue, which affects all subleading soft theorems, arises due to the long-range nature of gravitational interactions which lead to infrared divergences that render the standard S-matrix in four spacetime dimensions ill-defined.
The above power series expansion breaks down and receives non-analytic contributions in the form of logarithms \cite{Laddha:2018myi}. 

\paragraph{Logarithmic soft theorems.}
Given the existence of a well-defined soft expansion to all orders in the loop expansion, it was shown by Saha, Sahoo and Sen in a series of papers \cite{Sahoo:2018lxl,Saha:2019tub,Sahoo:2020ryf} that the ratio between an $(N+1)$-point amplitude with a soft graviton and the $N$-point amplitude without that soft particle is infrared-finite
with its soft expansion given by \cite{Sahoo:2018lxl}
\begin{equation}\label{logsoftexp}
    \frac{{\cal M}_{N+1}(p_{1}, \dots, p_{N};(\omega,q,\ell))}{{\cal M}_{N}(p_{1}, \dots, p_{N})}=
    \sum_{n=-1}^{\infty} \omega^{n} (\ln\omega)^{n+1} {S}^{(\ln \omega)}_{n}+\dots\,.
\end{equation}
The $n=-1$ term is the leading (Weinberg) soft factor ${S}^{(\ln \omega)}_{-1}\equiv {S}_{-1}$. %
For $n\ge0$ the soft expansion differs from the tree-level one \cite{Cachazo:2014fwa}.
In seminal work by Sahoo and Sen \cite{Sahoo:2018lxl} the leading logarithmic soft factor $S^{(\ln \omega)}_0\neq S_0$ was shown to be universal and uniquely fixed by the momenta of the scattering states, and in \cite{Sahoo:2020ryf} this was shown to be true also for $S^{(\ln \omega)}_1\neq S_{1}$. 
Moreover, it is conjectured \cite{Saha:2019tub} 
that the all-order soft expansion contains a universal tower of soft factors $S^{(\ln \omega)}_n$ (the $\dots$ contain non-universal terms of the form $\omega^n (\ln \omega)^m$ with $m\neq n+1$).
This would mean that the logarithmic soft theorems are valid to all orders in perturbation theory and independent of the details of the hard scattering! 
If this is indeed true, we expect to uncover asymptotic symmetries whose associated conservation laws give rise to the logarithmic soft graviton theorems. 

Our focus in this work is the (leading) logarithmic soft theorem of Einstein gravity coupled to matter. The log soft factor may be split as \cite{Sahoo:2018lxl}
\begin{equation}
\label{Sln0}
    {S}^{(\ln \omega)}_{0}={S}^{(\ln \omega)}_{0,{\rm classical}}+\Delta{S}^{(\ln \omega)}_{0, {\rm quantum}}.
\end{equation}
The classical log soft factor ${S}^{(\ln \omega)}_{0,{\rm classical}}$ can be derived from purely classical scattering and is valid universally, i.e. it is independent of the theory and the nature of the external particles \cite{Saha:2019tub, Sahoo:2021ctw}.
Alternatively, it can be derived from a quantum amplitude computation which comes with an additional contribution $\Delta{S}^{(\ln \omega)}_{0, {\rm quantum}}$.
This quantum log soft factor is also expected to be universal but has been obtained from direct one-loop computations in Einstein gravity \cite{Sahoo:2018lxl} (see also \cite{Bern:2014oka}).
Neither of the contributions to \eqref{Sln0} turns out to contain factors of $\hbar$  and so the split into classical and quantum is only possible thanks to the existence of a classical scattering computation \cite{Laddha:2018myi,Sahoo:2018lxl,Saha:2019tub} which yields the classical log soft theorem.

In Einstein gravity the classical log soft graviton factor is \cite{Sahoo:2018lxl}
\begin{equation}
\badat{2}
\label{Sln0Gclassical}
    {S}^{(\ln \omega)}_{0,{\rm classical}}&=\frac{i(\frac{\kappa}{2})^3}{8\pi} \sum_{i=1}^N  \frac{\varepsilon_{\mu\nu} p_i^\nu  q_\rho }{ q\cdot p_i}\sum_{\stackrel{j\neq i}{\eta_i \eta_j=1}} \frac{(p_i \cdot p_j)\left[p^\mu_i p^\rho_j-p^\mu_j p^\rho_i\right]\left[2(p_j\cdot p_j)^2-3p_i^2 p_j^2\right]}{\left[(p_i \cdot p_j)^2-p_i^2 p_j^2\right]^{3/2}}\\
    &\quad -\frac{i(\frac{\kappa}{2})^3}{4\pi} \,\sum_{i=1}^N \frac{\varepsilon_{\mu\nu}p_i^\mu p_i^\nu}{q\cdot p_i}\sum_{j,\eta_j=-1} ( q\cdot p_j).
\eadat
\end{equation}
The first term results from the late time gravitational radiation due to the late time acceleration of the particles via long range gravitational interaction. 
It arises in the quantum computation from the region where the loop momentum is large compared to the soft energy~$\omega$ but small compared to the energies of the other particles. 
The second term represents the effect of gravitational drag on the soft graviton due to the other finite energy particles in the final state which results in a time delay for the soft graviton to travel to the detector. In the quantum computation this term originates from loop momenta smaller than $\omega$ and larger than the infrared cut-off given by the inverse distance to the detector.\footnote{The drag also contributes to the $O(\w^0)$ term in the soft expansion, with a term proportional to $\ln R$ where $R$ is the largest scale in the problem given by the distance to the detector.}

In the quantum computation there are additional terms
\begin{align}
\label{Sln0sGquantum}
\badat{2}
   \Delta {S}^{(\ln \omega)}_{0, {\rm quantum}}&=-\frac{(\frac{\kappa}{2})^3}{16\pi^2} \sum_{i=1}^N\frac{\varepsilon_{\mu\rho} p_i^\rho  q_\nu}{ q\cdot p_i}\left(p^\mu_i\partial^\nu_{p_i}-p^\nu_i\partial^\mu_{p_i}\right)\\
    &\qquad
    \sum_{j\neq i}  \frac{2(p_i\cdot p_j)^2-p_i^2 p_j^2}{\sqrt{(p_i\cdot p_j)^2-p_i^2 p_j^2)}}\ln\left(\frac{p_i\cdot p_j+\sqrt{(p_i\cdot p_j)^2-p_i^2p_j^2}}{p_i\cdot p_j-\sqrt{(p_i\cdot p_j)^2-p_i^2p_j^2}}\right)\\
    &\quad - \frac{(\frac{\kappa}{2})^3}{8\pi^2}\,\sum_{i=1}^N \frac{\varepsilon_{\mu\nu}p_i^\mu p_i^\nu}{q\cdot p_i}\sum_{j,\eta_j=-1}  (\hat q\cdot p_j)\log \frac{m_j^2}{( q \cdot p_j)^2},
\eadat
\end{align}
which originate again from different regions of loop momentum integration: the first arises from the region where the loop momentum is large compared to the soft energy $\omega$ but small compared to the energies of the other particles, while the second term originates from loop momenta smaller than $\omega$ and larger than the infrared cut-off given by the inverse distance to the detector. %

In this work we will focus on deriving a symmetry interpretation for the classical log soft graviton factor \eqref{Sln0Gclassical} while quantum infrared effects %
will be discussed elsewhere \cite{CLPquantumlog}.\footnote{An earlier study of the symmetry underlying the loop-corrected subleading soft graviton theorem was carried out in \cite{Agrawal:2023zea}, but there is a discrepancy between their conservation law and the (classical and quantum) log soft theorem of Sahoo-Sen \cite{Sahoo:2018lxl} which was argued to be due to error in \cite{Sahoo:2018lxl}. 
However, in our understanding the results of \cite{Sahoo:2018lxl} are correct: our first principles derivation of the conservation law for superrotation symmetry is in perfect agreement with the (classical) log soft graviton theorem of \cite{Sahoo:2018lxl, Saha:2019tub}; we expect this equivalence to continue to hold in the quantum theory. Moreover, our analysis allows us to directly compute the tail to the displacement memory (a classical observable) which is unclear how to extract from~\cite{Agrawal:2023zea}.}

\paragraph{Conservation laws for superrotations.}

Our goal is to give a first-principles derivation of the classical logarithmic soft graviton theorem from infrared-finite conservation laws in Einstein gravity coupled to massive matter. We will show that these conservation laws arise from superrotation symmetry.
The starting point is the computation of the symplectic structure at the asymptotic boundary 
\begin{align}
\label{Omegapm}
  \Omega_{ i^{\pm}\cup {\cal I}^{\pm}}=\Omega^{\rm mat}_{i^{\pm}}+\Omega^{\rm rad}_{{\cal I}^{\pm}},
\end{align}
which consists of a matter contribution at $i^{\pm}$ and a radiative contribution at ${\cal I}^{\pm}$.
In the absence of long-range infrared effects it has been shown \cite{Kapec:2014opa,Campiglia:2014yka,Campiglia:2015kxa,Campiglia:2015yka} that one can construct charges from the symplectic structure
\begin{equation}
\Omega_{\,i^\pm\,\cup \,\cal I^\pm\,}(\delta,\delta_Y)=\delta Q^{\pm},
\end{equation}
when one of the field variations, denoted here by $\delta_Y$, implements superrotation symmetry. %
Their classical conservation law $Q^+=Q^-$,
elevated at the level of the S-matrix to the commutator
\begin{equation}
    \langle {\rm out}|Q^+ \mathcal S-\mathcal SQ^-|{\rm in}\rangle=0,
\end{equation}
is equivalent to the subleading tree-level soft graviton theorem \cite{Kapec:2014opa,Campiglia:2014yka,Campiglia:2015kxa}.
In this work we revisit this connection when classical infrared effects are accounted for.

Due to the long-range nature of gravitational interactions asymptotic matter fields are not free but `dressed' by a phase \cite{Kulish:1970ut},
\begin{equation}
\label{phidressed}
    \varphi=e^{i { \Phi}} \varphi_{ \rm free}^++e^{-i { \Phi}} \varphi_{\rm free}^-,
\end{equation}
where $ \varphi_{\rm free}^\pm$ denotes the positive/negative frequency modes of the real massive scalar field~$\varphi$.
As we will show in section \ref{sec:AsymptoticsGravity}, the dressing ${ \Phi}$ comes in the form of logarithms at early and late times and is tied, via the equations of motion, to the asymptotic conditions of the metric in the form of so-called ``tails''.
These are weaker compared to the standard radiative fall-offs in the absence of long-range infrared effects. In section \ref{sec:superrotation} we will identify superrotations that are consistent with these asymptotics and smoothly interpolate between $i^\pm$ and $\scri^\pm$. 

The asymptotic behavior of the matter and radiation fields together with the superrotations defined on $i^\pm \cup \scri^\pm$ serve as input to computing the IR corrected symplectic structure which differs from its free counterpart. 
The matter dressing and the radiation tails cause the symplectic structure to diverge at early and late times, both on the time-like and null boundaries, as we will show in section \ref{sec:SymplecticStructure}. %
To regulate these IR divergences we introduce a late-time cutoff $\Lambda^{-1}$ for both boundaries. 
We then show that for superrotations the symplectic structure can be expressed as a total variation on field space.
This allows us to construct a regularized Noether charge
\begin{equation}
\label{NoetherQ}
     Q^{\Lambda}=
        \ln \Lambda^{-1}\left(Q^{(\ln)}_H+ Q^{(\ln)}_S \right) + \left(Q^{(0)}_H+ Q^{(0)}_S\right)+ \dots,
\end{equation}
where the subscripts $H$ and $S$ refer to the `hard' and `soft' contributions to the total charge which originate from, respectively, the $i^\pm$ and the $\scri^\pm$ contributions to \eqref{Omegapm}. %
In section \ref{sec:Gravitycharges} we discuss how the conservation of this charge is connected to soft graviton theorems.

The finite charge 
\begin{equation}
\label{Q0}
    Q^{(0)}\equiv Q^{(0)}_H+ Q^{(0)}_S,
\end{equation} 
whose explicit expression will be given in \eqref{GravityQH0+} and \eqref{GravityQS0+}, in fact matches the one derived in \cite{Campiglia:2015kxa} where infrared effects were ignored. It obeys a classical conservation law which in an S-matrix element becomes the subleading tree-level soft graviton theorem
\begin{equation}
  Q^{(0)}_+=Q^{(0)}_-\quad  \Longleftrightarrow \quad \lim_{\omega\to 0}(1+\omega \partial_\omega) {\cal M}_{N+1}=
    S_{0} {\cal M}_N.
\end{equation}
However, this subleading soft theorem, which behaves as $\omega^0$ as $\omega \to 0$, becomes ambiguous due to the presence of infrared effects. This is also visible at the level of the Noether charge \eqref{NoetherQ} where any rescaling of the IR cutoff $\Lambda$ renders any $\Lambda^0$ term ambiguous. 

Instead, the log charge
\begin{equation}
\label{Qln}
 Q^{(\ln)}\equiv    Q^{(\ln)}_H + Q^{(\ln)}_S
\end{equation}
is unambiguous, and its explicit expression will be given in \eqref{GravityQHln+} and \eqref{GravityQSln+}. Because it is also conserved we can divide both sides by $\ln \Lambda^{-1}$ and subsequently take $\Lambda \to 0$. In section~\ref{sec:Gravitylogcharges} we will prove that the resulting conservation law yields the classical logarithmic soft theorem
\begin{equation}
    Q^{(\ln)}_+=Q^{(\ln)}_-\quad  \Longleftrightarrow \quad \lim_{\omega \to 0}\partial_\omega \omega^2 \partial_\omega {\cal M}_{N+1}=
    S^{(\ln \omega)}_{0,{\rm classical}} {\cal M}_N.
\end{equation}
In \cite{Choi:2024ygx}, and in much more detail here, we provide a first-principles covariant phase space derivation of this charge. Moreover, we will extend our earlier results by proving that the logarithmic superrotation charge \eqref{Qln} is exact to all orders in the gravitational coupling~$\kappa$ which beautifully echoes the one-loop exactness of the logarithmic soft graviton theorem.

\bigskip
This paper is organized as follows. In section \ref{sec:Preliminaries} we collect the basics for discussing the physics near null and time-like infinity. To compute superrotation charges in the covariant phase space formalism that account for infrared effects we need two main ingredients: 1) the asymptotic data of the gravity and matter fields  in the presence of long-range interactions which we derive in section \ref{sec:AsymptoticsGravity} and from which we extract the displacement memory and its tail; and 2) the superrotation symmetry transformation which, as we show in section \ref{sec:superrotation}, smoothly interpolates across the time-like and null boundaries. In section \ref{sec:SymplecticStructure} we compute the symplectic structure, regulate it and extract the infrared corrected superrotation charges. Finally we proof in section \ref{sec:Gravitycharges} that the conservation law associated to our novel superrotation charge is in fact the classical logarithmic soft graviton theorem. 
In Appendices \ref{app:UsefulFormulas}, \ref{app:tails} and \ref{app:Green} we collect various formulas necessary for this derivation, and
in Appendix \ref{app:Gravityallorder} we show that our novel superrotation charge conservation law is exact to all orders in the gravitational coupling $\kappa$.
\section{Preliminaries}
\label{sec:Preliminaries}
The symplectic structure $\Omega$ is defined on a Cauchy slice $\Sigma$ and we have to take great care in pushing it to the past and future boundary, which each contains a time-like component $i^\pm$ and a null component $\cal I^\pm$. Starting from (asymptotically) Minkowski space %
\begin{equation}
    ds^2=\eta_{\mu\nu}dx^\mu dx^\mu=-dt^2+d\vec x\cdot d\vec x,
\end{equation}
where $x^\mu=(t,\vec x)=(t,r \hat x)
$ with $r^2=\vec{x}\cdot \vec{x}$,
we briefly review the limits that land us on~\eqref{Omegapm}. See also Appendix \ref{app:UsefulFormulas}.

\paragraph{Radiation.}
Massless particles such as gravitons reach future null infinity, $\scri^+ \simeq S^2 \times \mathbb R$, in the limit $t + r \to \infty$ at $t - r = {\rm fixed}$, and so we parametrize it by retarded Bondi coordinates $u = t - r$ in which the Minkowski metric takes the form
\begin{equation}
\label{ds2Bondi}
    ds^2=-du^2-2du dr+r^2\gamma_{AB} dx^A dx^B,
\end{equation}
with $x^A=(z,\bar z)$ denoting the angles and $\gamma_{AB}$ the round metric on the two-sphere.  A spacetime point in Bondi coordinates is then written as\footnote{
    We use the notation $\hat x$ to denote a 3-vector restricted to the unit sphere.
    In a slight abuse of notation, we also use $\hat x$ to denote the angular variables: $d^2\hat x\equiv dzd\zb\gamma_{z\zb}$ is the sphere volume element, and $f(\hat x)$ denotes a function on the sphere.
}

\begin{equation}
    x^\mu=u \,t^\mu+r \,q^\mu \quad \text{with} \quad t^\mu=(1,\vec0),\; q^\mu=(1,\hat x)
\end{equation}
The radiative contribution to the symplectic structure \eqref{Omegapm} is then computed by taking the $t$~=~constant time slice $\Sigma_t$, on which it is defined, to the future boundary  
\begin{equation}
\label{Omrad}   
  \Omega^{\rm rad}_{{\cal I}^{+}}=      \lim_{\Sigma_t\to \scri^+} \Omega_{t}^{\rm rad}.
\end{equation}
The corresponding contribution on the past boundary is computed in an analogous way in terms of the advanced Bondi coordinate $v = t + r$ which is held fixed as $t-r\to -\infty$. These limits have to be taken carefully and we will discuss important subtleties below.
\paragraph{Matter.}
Massive particles originate from past time-like infinity $i^-$ and end up on future time-like infinity $i^+$ which are reached by taking $t\to \mp \infty$ while holding $r/t$ = fixed. To describe the neighbourhood of $i^\pm$ we use hyperbolic, or Euclidean AdS$_3$, coordinates which cover the interior of the past and future light cones with the vertex at an arbitrary point.
The one-parameter family of coordinates inside the future light cone of the point $(u_0,\vec{0})$ %
is defined by
\begin{equation}
\label{taurho}
\tau = \sqrt{t^{2} - r^{2}} - u_{0}
,\qquad \tau \rho = r.
\end{equation}
It can be checked that any fixed $\tau$ surface intersects ${\cal I}^{+}\simeq S^2 \times \mathbb R$ at $u = u_{0}$ in the limit $\rho\to \infty$. Throughout this paper we will fix $u_{0} = 0$ as in \cite{Campiglia:2015qka}, which implies that the $(\tau,\rho,x^A)$ coordinates cover the (interior) of the future light cone with tip at $(0,\vec{0})$. 
The Minkowski metric can be written as
\begin{equation}
\label{ds2H3}
       ds^2=
        -d\tau^2
        + \tau^2 k_{\alpha\beta} dy^\alpha dy^\beta,
\end{equation}
where $y^\alpha=(\rho,x^A)$ denote the coordinates on the three-dimensional space-like hyperboloid $\mathcal{H}_\tau$ at constant $\tau$ with metric 
\begin{equation}
    k_{\alpha\beta} dy^\alpha dy^\beta=
        \frac{d\rho^2}{1+\rho^2}
        + \rho^2\gamma_{AB}dx^Adx^B.
    \label{rhox}
\end{equation}
We will denote by
\begin{equation}\label{Y}
  \Y^\mu=(\sqrt{1+{\rho}^2},{\rho} \hat x)
\end{equation}
a unit vector on Minkowski space parametrized by $y^\alpha=(\rho,x^A)$ and
write a spacetime point in hyperbolic coordinates as
\begin{equation}
     {x}^\mu=\tau \Y^\mu.
\end{equation}
The `blow-up' of future time-like infinity $i^{+}$ is then given by $\mathcal H_\tau$ as $\tau\, \rightarrow\, \infty$ at fixed $\rho$. %
The matter contribution to the symplectic potential \eqref{Omegapm} on the future boundary is then\footnote{Note that the hyperboloid $\mathcal H_\tau$ is actually not a Cauchy slice but describes only its $t\geq r$ portion. The completion to a Cauchy slice requires the inclusion of the $t< r$ portion of future null infinity $\scri^+$ with which $\mathcal H_\tau$ intersects at $t=r$. Since $\phi(u,r,z^A)\stackrel{r\to \infty}{=}O(r^{-3/2})$ at fixed $u=t-r$ this null infinity contribution to the matter symplectic potential vanishes and \eqref{Ommat} is the full answer. See \cite{Campiglia:2015qka} for more details.}
\begin{equation}
\label{Ommat}
 \Omega^{\rm mat}_{i^{+}}=\lim_{{\cal H}_\tau\to i^+}\Omega_{\tau}^\text{mat}.
\end{equation}
In a similar way the matter contribution on the past boundary is computed, in coordinates $(\tau,\rho)$ inside the past light cone of the origin, by taking the limit $\tau \to -\infty$ at fixed $\rho$ which takes $\mathcal H_\tau \to i^-$. 
\\

Together the contributions \eqref{Omrad} and \eqref{Ommat} compute the symplectic structure on a Cauchy slice spanned by the union of the entire future boundary $i^{+}\, \cup\, {\cal I}^{+}$, for which we will derive explicit expressions in section \ref{sec:SymplecticStructure}. 
The celestial sphere on $\scri^+$ at $u=+\infty$ can be thought of as the $\rho\to \infty$ boundary of the asymptotic hyperboloid at $\tau=+\infty$. 
This will allow us to relate the asymptotic (late time) expansions of the matter and radiative fields in section \ref{sec:AsymptoticsGravity}, and enables us to write down a superrotation vector field that smoothly interpolates across time-like and null infinity in section \ref{sec:superrotation}. A similar analysis can be carried out for the past boundary $i^{-}\, \cup\, {\cal I}^{-}$. 
\section{Asymptotic data in gravity}
\label{sec:AsymptoticsGravity}

We now turn to the analysis of superrotation symmetries of gravitational scattering in four spacetime dimensions focusing on Einstein gravity with a minimally coupled massive real scalar field.  We start by discussing the asymptotic phase space. The field content are a massive real scalar field $\varphi$ coupled minimally to a metric $g_{\mu\nu}$ with action
\begin{align}
    S
    &=
        \int d^4x\sqrt{-g}\left(
            \frac{2}{\kappa^2}  R
            - \frac12 g^{\mu\nu}\p_\mu\varphi\p_\nu\varphi
            - \frac12 m^2\varphi^2
        \right)
    .
\end{align}
Our primary motivation is to analyse the asymptotic conservation laws for gravitational scattering which uses a perturbative expansion
in the coupling constant $\kappa = \sqrt{32\pi G_N}$ of the metric
\begin{align}
g_{\mu\nu} = \eta_{\mu\nu} + \kappa \,h_{\mu\nu}
\end{align}
around a fixed Minkowski vacuum spacetime $\eta_{\mu\nu}$ and perturbation $h_{\mu\nu}$. 
The indices of $h_{\mu\nu}$ are lowered and raised by the flat background metric $\eta_{\mu\nu}$ and its inverse. We will adopt this perturbative approach in the following, but we emphasize that our final result for the charges, whose conservation law will turn out to be associated to the logarithmic soft graviton theorem, are {\it exact to all orders in the coupling $\kappa$} as we will show in Appendix~\ref{app:Gravityallorder}.

We work in de Donder (harmonic) gauge,
\begin{align}\label{deDonder}
    \nabla^\nu h_{\mu\nu} - \frac12\nabla_\mu h = 0,
\end{align}
where %
$h=\eta^{\mu\nu} h_{\mu\nu}$ denotes the trace, and $\nabla$ denotes covariant derivative compatible with the flat background $\eta$.
The Einstein field equations are
\begin{equation}
\label{Einstein}
     R_{\mu\nu}- \frac12 g_{\mu\nu}  R
	= \frac{\kappa^2}{4} T_{\mu\nu},
\end{equation}
where the Einstein tensor $G_{\mu\nu}=R_{\mu\nu}- \frac12 g_{\mu\nu}  R$, in de Donder gauge and to linear order in~$\kappa$, is given by
\begin{equation}
    G_{\mu\nu}
    =-\frac\kappa 2\left(
    \nabla_\mu\nabla_\nu h
    - \nabla_\lambda \nabla_\mu h^\lambda_\nu
    - \nabla_\lambda \nabla_\nu h^\lambda_\mu
    + \nabla^2 h_{\mu\nu}
    - \frac12g_{\mu\nu}\nabla^2 h
    \right).
\end{equation}
The stress-energy tensor 
\begin{equation}
    T_{\mu\nu}=T^{\rm matt}_{\mu\nu}+T^h_{\mu\nu}
\end{equation}
receives contributions from matter and gravitons.
In de Donder gauge in a flat background the graviton stress tensor is given by\footnote{The (anti)symmetrization of indices are defined with unit weight, i.e.\ $t_{(\mu\nu)}\equiv\frac12(t_{\mu\nu}+t_{\nu\mu})$ and $t_{[\mu\nu]}\equiv\frac12(t_{\mu\nu}-t_{\nu\mu})$.}
\begin{equation}\label{Th}
\badat{2}
    T^h_{\mu\nu}
    &=
        2h^{\sigma\lambda}
            \left(
                2\nabla_\sigma\nabla_{(\mu} h_{\nu)\lambda}
                - \nabla_\mu\nabla_\nu h_{\sigma\lambda}
                - \nabla_\sigma\nabla_\lambda h_{\mu\nu}
            \right)
        - h_{\mu\nu}\nabla^2 h
        + 4\nabla^\sigma h_\mu^\lambda \nabla_{[\lambda} h_{\sigma]\nu}
        \\&\quad
        - \nabla_\mu h^{\sigma\lambda} \nabla_\nu h_{\sigma\lambda}
        + g_{\mu\nu} \left[
            2h_{\sigma\lambda}\nabla^2 h^{\sigma\lambda}
            + \nabla^\kappa h^{\sigma\lambda}\left(
                \frac32 \nabla_\kappa h_{\sigma\lambda}
                - \nabla_\sigma h_{\lambda\kappa}
            \right)
        \right]
    ,
    \eadat
\end{equation}
while the  matter stress-energy tensor is
\begin{equation}
\label{Tmattfree}
    T^{\rm matt}_{\mu\nu}=\nabla_\mu \varphi\nabla_\nu\varphi
        - \frac12 \eta_{\mu\nu}(\eta^{\lambda\sigma}\nabla_\lambda\varphi\nabla_\sigma\varphi + m^2\varphi^2).
\end{equation}
In de Donder gauge the scalar equation of motion to linear order in $\kappa$ is given by
\begin{align}\label{eom_gr_realscalar}
	\left[\nabla^2
	-m^2
	- \kappa h_{\mu\nu}
		\nabla^\mu\nabla^\nu\right]\varphi
	&=
		0
	.
\end{align}

In turn we will discuss the asymptotic fall-offs for the real massive scalar matter and the metric. We will focus on the asymptotics at $i^+\cup \scri^+$, but a similar analysis can be repeated at $i^-\cup \scri^-$.

\subsection{Matter %
}

We start with the discussion of the late-time behavior of a real massive scalar field minimally coupled to gravity. We use hyperbolic coordinates $(\tau, y^\alpha$) = $(\tau, \rho, x^A)$ in terms of which the de Donder condition, Einstein equations and the equation of motion for the scalar field are given in Appendix~\ref{app:UsefulFormulas}.
In what follows we will derive the late-time expressions of the matter and graviton fields to all orders in the coupling $\kappa$. We will see that the interaction of the free matter field with the graviton field leads to a {\it logarithmic dressing} of the matter field and a {\it tail} at late times of the gravitational field.

The asymptotics of a {\it free} real massive scalar field in Minkowski space,
\begin{equation}
\label{freescalar}
    \varphi_\text{free}(x)=\int \frac{d^3\vec{p}}{(2\pi)^3 2E_p} \left[b(\vec{p}) e^{ip\cdot x}+h.c.\right]
\end{equation}
with $E_p=\sqrt{|\vec{p}|^2+m^2}$ and $p\cdot x=-E_pt+\vec{p}\cdot \vec{x}$, is described along time-like geodesics and its late-time ($\tau \to \infty$) behavior can be extracted via a saddle point approximation with critical point $\vec{p}=m\rho\hat{x}$%
.
After absorbing various %
phases into the `free data' given by $b$, the late-time asymptotics of the free scalar field can be expressed as 
\begin{equation}
    \label{varphi0}
    \varphi_\text{free}(\tau, y) \stackrel{\tau\to\infty}{=} 
    \frac{\sqrt m}{2(2\pi\tau)^{\frac{3}{2}}}
    b(y) e^{- i m \tau} + {\rm c.c.}\,.
\end{equation}
Via Einstein's equations the free matter stress tensor \eqref{Tmattfree}
sources a `Coulombic' potential.
The $\tau\tau$ component of Einstein's equations \eqref{Einstein} is of order $O(1/\tau^3)$ and sources a $1/\tau$ late time `tail' in the metric
\begin{align}\label{httfalloff}
    h_{\tau\tau}(\tau,y)
    \overset{\tau\to\infty}{=}\frac1\tau \overset1h_{\tau\tau}(y)+\dots
\end{align}
where 
\begin{equation}\label{h1tt_diffeq}
    (\calD^2-3)\overset1h_{\tau\tau}
            =-\frac{\kappa}{4}\overset{3}{T}{}^{\rm matt}_{\tau\tau}       .
\end{equation}
The de Donder gauge condition \eqref{deDondertaualpha} then implies
\begin{equation}\label{htahabfalloff}
\badat{2}
   h_{\tau\alpha}(\tau,y)&\overset{\tau\to\infty}{=}
        \overset0h_{\tau\alpha}(y) + \dots
   ,\\
   h_{\alpha\beta}(\tau,y)&\overset{\tau\to\infty}{=}
        \tau \overset{-1}h_{\alpha\beta}(y) + \dots
\eadat
\end{equation}
where %
\begin{equation}
\badat{2}
    \frac15\left(
            2{\cal D}^\alpha \overset0h_{\alpha\tau}
            - k^{\alpha\beta} \overset{-1}h_{\alpha\beta}
        \right)&=\overset1h_{\tau\tau}
    ,\\
    \frac13\left(
            {\cal D}^\beta \overset{-1}h_{\beta\alpha}
            - \frac12 k^{\beta\gamma}{\cal D}_\alpha \overset{-1}h_{\beta\gamma}
            + \frac12 {\cal D}_\alpha \overset1h_{\tau\tau}
        \right)    &=
        \overset0h_{\tau\alpha}    .
\eadat
\end{equation}
These `Coulombic' modes all arise from the interactions and come with one power of the coupling constant, $\overset1h_{\tau\tau}=O(\kappa)$, $\overset1h_{\tau\alpha}=O(\kappa)$ and $\overset1h_{\alpha\beta}=O(\kappa)$. 
The scalar equation of motion~\eqref{eomvarphi}, with  $(\nabla^2-m^2-\kappa \overset1h_{\tau \tau}\tau^{-1}\partial_\tau^2+\dots)\varphi=0$ the leading correction to the free equation of motion, is solved at late times by\footnote{In Appendix~\ref{app:Gravityallorder} we show that higher powers of $\ln \tau$ which would contribute to \eqref{varphiati+} at order $O(\kappa^3$) are in fact absent and the result \eqref{varphiati+}-\eqref{Gravity_expansion} is thus exact to all orders in the coupling $\kappa$.}
\begin{equation}
\label{varphiati+}
    \varphi(\tau,y)\overset{\tau\to\infty}{=}
        \frac{e^{-im\tau}}{\tau^{3/2}}
        \left(
            \ln\tau
            \overset\ln b_0(y)
            + b_0(y)
           + \dots
        \right)
        + \text{c.c.},
\end{equation}
where $b_0 \equiv \sqrt{\frac{m}{4(2\pi)^3}}\,b$ is the free asymptotic data, and
\begin{equation}\label{Gravity_expansion}
    \overset\ln b{}_0(y)
    =
        \frac{i\kappa}2m \overset1h_{\tau\tau}(y)b_0(y).
\end{equation}
This expression is {\it exact in the coupling $\kappa$}.

The late-time behavior of the {\it dressed} matter field \eqref{varphiati+} at $i^+$ determines, via Einstein's equations \eqref{Einsteinhyperbolic}, the asymptotic form of the matter stress tensor
\begin{equation}\label{Tmattfalloff}
\badat{3}
    T_{\tau \tau}^{\rm matt}(\tau,y)&\overset{\tau\to\infty}{=}
        \frac{1}{\tau^3}\overset3T{}_{\tau \tau}^{\rm matt}
        + \frac{\ln\tau}{\tau^4}\overset{4,\ln}T{}_{\tau \tau}^{\rm matt}
        + \frac{1}{\tau^4}\overset4T{}_{\tau \tau}^{\rm matt}
        + O\Big(\frac{\ln\tau}{\tau^5}\Big)
    \\
    T_{\tau \alpha}^{\rm matt}(\tau,y)&\overset{\tau\to\infty}{=}
       \frac{\ln\tau}{\tau^3}\overset{3,\ln}T{}_{\tau \alpha}^{\rm matt}+\frac{1}{\tau^3}
       \overset3T{}_{\tau\alpha}^{\rm matt}+O\Big(\frac{\ln\tau}{\tau^4}\Big)
    \\
    T_{\alpha \beta}^{\rm matt}(\tau,y)&\overset{\tau\to\infty}{=}
        O\Big(\frac{\ln\tau}{\tau^3}\Big).
\eadat
\end{equation}
While $T_{\tau \tau}^{\rm matt}$ at leading order is given by the free matter stress tensor, $T_{\tau \alpha}^{\rm matt}$ receives a logarithmic corrections that dominates over the free expression.\footnote{Naively, also $T_{\tau \tau}^{\rm matt}$ receives a leading logarithmic correction but its coefficient vanishes. Higher powers of $\ln \tau$ are again absent - see Appendix \ref{app:Gravityallorder}.}
The explicit expressions for the relevant stress tensor components are \begin{equation}\label{Tmatt}
    \overset3T{}_{\tau \tau}^{\rm matt}=2m^2b_0^* b_0
    ,\quad
    \overset3T{}_{\tau\alpha}^{\rm matt}=im (b_0^*\p_\alpha b_0-b_0\p_\alpha b_0^*)
    ,\quad
    \overset{3,\ln}T{}_{\tau \alpha}^{\rm matt}=- \kappa m^2(\p_\alpha\overset1h_{\tau\tau}) b_0^* b_0
    .
\end{equation}
Meanwhile, the late-time behavior of gravitons at $i^+$ determines the asymptotic form of the graviton stress tensor,
\begin{align}\label{Thfalloff}
    T^h_{\tau\tau}
    &=
        \frac{1}{\tau^4}\overset4T{}^h_{\tau\tau}
        + \cdots
    ,\qquad
    T^h_{\tau\alpha}
    =
        \frac1{\tau^3}\overset3T{}^h_{\tau\alpha}
        + \cdots
    ,\qquad
    T^h_{\alpha\beta}
    =
        \frac1{\tau^2}\overset2T{}^h_{\alpha\beta}
        + \cdots
    .
\end{align}
Notice that the $\tau\tau$ component of the graviton stress tensor is subleading compared to that of the matter stress tensor, while the $\tau\alpha$ component is at the same order as the subleading matter stress tensor component. We will see how this interplay enters the superrotation charges below.

\subsection{Radiation %
}
At future (past) null infinity $\scri^+$ ($\scri^-$) we use retarded (advanced) Bondi coordinates $(u\,(v), r, x^A)$ in terms of which the de Donder condition and Einstein equations are given in Appendix~\ref{app:UsefulFormulas}.
We study the asymptotics of the radiative graviton field at early/late times $u\to \pm \infty$ ($v\to \pm\infty$); we focus again on the future boundary. 

The asymptotics of the {\it free} gravitational field in Minkowski space,
\begin{equation}
    h^{\rm free}_{\mu\nu}(x)=
    \int \frac{d^3k}{(2\pi)^32\omega_k}\left[a_{\mu\nu}(\vec k)e^{ik\cdot x}+{a}_{\mu\nu}(\vec k)^\dagger e^{-ik\cdot x}\right]
\end{equation}
with $\omega_k=k^0=|\vec k|$, $k\cdot x= -\omega_k t+\vec k \cdot \vec x$ and $a_{\mu\nu}(\vec k)=\sum_{\lambda=\pm} \varepsilon^{\lambda*}_{\mu\nu} a_\lambda(\vec k)$, can be obtained from a saddle point analysis with saddle point $\vec k=\omega_k\hat{x}$%
. The $r\to \infty$ limit (at fixed $u$) is%
\begin{equation}
    h^{\rm free}_{\mu\nu}(x)=-\frac{i}{8\pi^2 r}\int_0^\infty d\omega_k \left[a_{\mu\nu}(\omega_k,\hat x)e^{-i\omega_k u}-a_{\mu\nu}(\omega_k,\hat x)^\dagger e^{i\omega_k u}\right]+O(1/r^2).
\end{equation}
The free data is given by the two independent polarizations of the graviton. Without loss of generality we write the polarization tensors as $\varepsilon_{\mu\nu}^\pm=\varepsilon^\pm_\mu \varepsilon^\pm_\nu$ where
\begin{equation}
    \varepsilon_\mu^+=\frac{1}{\sqrt{2}}(-\bar z, 1, -i, -\bar z),\quad  \varepsilon_\mu^-=\frac{1}{\sqrt{2}}(- z, 1, i, - z).
\end{equation}
After mapping the Cartesian components to Bondi coordinates we have
\begin{equation}
    \varepsilon^+_u=-\frac{\bar z}{\sqrt{2}},\quad \varepsilon^+_r=0,\quad \varepsilon^+_z=0, \quad \varepsilon^+_{\bar z}=\frac{\sqrt{2}r}{1+z\bar z},
\end{equation}
and similar expressions for the opposite helicity. This yields the large-$r$ behavior of the free graviton field 
\begin{equation}
    h^{\rm free}_{uu}=O(r^{-1}),\quad h^{\rm free}_{uA}=O(r^0),\quad 
   h^{\rm free}_{ur}=h^{\rm free}_{rr}=h^{\rm free}_{rA}=0 , \quad h^{\rm free}_{AB}=O(r).
\end{equation}

What about interactions?
In four spacetime dimensions, generic gravitational scattering leads to an asymptotic radiative gravitational field at ${\cal I}^{+}$ that violates the {\it peeling property} \cite{Christodoulou2002, Kehrberger:2021uvf, Bieri:2023cyn}: in the large-$r$ expansion it contains logarithmic terms that fall off as $O(\frac{\ln r}{r})$. The resulting spacetimes are known as asymptotically logarithmically flat spacetimes, \cite{Winicour1985, Geiller:2024ryw}. Fortunately, it was shown in \cite{Geiller:2024ryw} that superrotations continue to be asymptotic symmetries of asymptotically logarithmically flat spacetimes and the $\ln r$ terms in the Bondi expansion do not affect the final superrotation charge at ${\cal I}^{+}$ derived in \cite{Choi:2024ygx}. We can thus omit these $\ln r$ terms for the purpose of our analysis. 

Nonetheless, as the matter stress tensor at $i^{\pm}$ have logarithmic tails in $\tau$, it is pertinent to analyse the effect of these tails on the outgoing radiation as $u\to \pm \infty$. In \cite{Laddha:2018myi,Laddha:2018vbn,Saha:2019tub} it was shown that when the gravitational radiation is sourced by the stress tensor of massive point particles, the asymptotic behaviour of the radiative field $h_{\mu\nu}(u,\hat{x})$ as $u \rightarrow\pm \infty$  generates a so-called {\it tail to the memory}. Our analysis in this section shows that the tail to the memory persists even when the source is a smooth massive scalar field.
The late-time fall-off at the future boundary of $\scri^+$ can be inferred from the analysis at $i^+$. 

We start with the linearized Einstein equations in de Donder gauge sourced by the matter stress tensor, $\Box \bar h_{\mu\nu} = -\frac{\kappa}{2}T_{\mu\nu}$, which we expressed in terms of the trace-reversed metric 
$\bar h_{\mu\nu} = h_{\mu\nu} - \frac12 h\,\eta_{\mu\nu}$.
In Cartesian coordinates, its retarded solution is given by 
\begin{equation}
\label{hretarded} 
\bar h_{\mu\nu}(x)=\frac{\kappa}{4\pi}\int d^4x' \Theta(t-t')\delta\Big((x-x')^2\Big)T_{\mu\nu}(x'),
\end{equation}
which we evaluate at $\scri^+_+$ by first taking $r\to \infty$ at fixed $u$ and then letting $u\to \infty$ where the graviton field is sourced by the late-time asymptotics of the matter stress tensor at $i^+$. The above equation is premised on the fact that at the co-dimension two boundary of spacetime, $i^{+}\, \cap\, {\cal I}^{+}$, the co-ordinate system in the neighbourhood of $i^{+}$ can be smoothly glued with the retarded Bondi co-ordinates at $\cal I^+$. This has been shown in \cite{Compere:2023qoa}.

To that end we express the spacetime point $x^\mu$ in Bondi coordinates while we write ${x'}^\mu$ in hyperbolic coordinates using the expressions in section \ref{sec:Preliminaries}.
At large $r$ and fixed $u$ we have
\begin{equation}
    -(x-x')^2=2r(u+\tau' q\cdot \Y')+O(r^0),
\end{equation}
where $q^\mu=(1,\hat x)$. Using $d^4x'=d^3y' d\tau' {\tau'}^3$ we have to evaluate
\begin{equation}
\label{hretardedfin} 
\bar h_{\mu\nu}(x)=\frac{\kappa}{8\pi r}\int d^3y' d\tau'{\tau'}^3 \delta(u+\tau' q\cdot \Y')T_{\mu\nu}(x').
\end{equation}
The Cartesian components of the stress-energy tensor are related to the hyperbolic ones by
\begin{equation}\label{TmnCartesian}
    T_{\mu\nu}
    =
		\Y_\mu \Y_\nu T_{\tau\tau}
		- \frac1\tau \calD^\alpha(\Y_\mu \Y_\nu) T_{\tau\alpha}
		+ \frac{1}{\tau^2}(\calD^\alpha \Y_\mu) (\calD^\beta \Y_\nu) T_{\alpha\beta}
    .
\end{equation}
Inserting the large-$\tau$ behavior of the matter and graviton stress-energy tensors \eqref{Tmattfalloff} and \eqref{Thfalloff} at $i^+$, we can extract the large-$u$ behavior of the Cartesian components of the graviton field at $\scri^+$ as follows\footnote{We will show in Appendix \ref{app:Gravityallorder} that there are no higher powers of logarithms in the asymptotics of the graviton field \eqref{hmunu}.}
\begin{align}\label{hmunu}
	 h_{\mu\nu}(x)
	&=
		\frac1r\left[
			\overset0h_{\mu\nu}(x^A)
			+ \frac{\ln u}{u} \overset{1,\ln}{h_{\mu\nu}}(x^A)
			+ \frac1u \overset1h_{\mu\nu}(x^A)
            + \cdots
		\right],
\end{align}
where we used the fact that the metric and its trace-reversed form only differ at subleading order in $r$. 
Using energy-momentum conservation $\nabla^\mu T_{\mu\nu}=0$ one finds (see Appendix \ref{app:tails}) that the coefficient of the $\ln u/u$ term vanishes, while the constant term is sourced by the free matter energy-momentum tensor 
\begin{equation}\label{hmn0T}
    \overset0h_{\mu\nu}(x^A)=\frac{\kappa}{8\pi}\int d^3y (-q\cdot \Y)^{-1}\left(\Y_\mu \Y_\nu +\frac12\eta_{\mu\nu}\right)\overset3T{}^{\rm matt}_{\tau\tau},
\end{equation}
and the $1/u$ term is determined by the leading correction to the matter energy-momentum tensor due to long-range interactions and the graviton stress tensor
\begin{equation}
\badat{2}
    \overset1h_{\mu\nu}(x^A)
    &=
        -\frac{\kappa}{8\pi }\int d^3y\bigg[
            \frac{
                (q\cdot\Y)\calD^\alpha(\Y_\mu \Y_\nu)
                - (\Y_\mu \Y_\nu+\frac12\eta_{\mu\nu}) \calD^\alpha (q\cdot\Y)
            }{q\cdot \Y}
            \overset{3,\ln}{T}{}^{\rm matt}_{\tau\alpha}
            \\
            &- \left(
                (\Y_\mu \Y_\nu+\frac12\eta_{\mu\nu}) k^{\alpha\beta}
                + (\calD^\alpha\Y_\mu)(\calD^\beta\Y_\nu)
                - \frac12\eta_{\mu\nu}(\calD^\alpha\Y_\sigma)(\calD^\beta\Y^\sigma)
            \right)\overset2T{}^h_{\alpha\beta}
        \bigg]
    .
\eadat
\end{equation}
The relevant Bondi components are $r C_{AB}=(\partial_A x^\mu)(\partial_B x^\nu) \kappa h_{\mu\nu}$.

Let us summarize these results for the radiation field at null infinity.
The free data of the graviton field at $\scri^+$ in retarded Bondi coordinates where it solves Einstein's equations \eqref{hbar}-\eqref{Einsteinretarded} is given by 
\begin{equation}
\label{hABlarger}
    \kappa h_{AB}(u,r,x^A)\stackrel{r\to \infty}{=} r C_{AB}(u,x^A)+\dots\,,
\end{equation}
while the de Donder gauge condition \eqref{deDonderretarded}  implies
\begin{equation}
\label{standardhlarger}
 h_{uu}=O(r^{-1}),\quad h_{uA}=O(r^0),\quad h_{ur}=h_{rr}=h_{rA}=O(r^{-2}).
\end{equation}
At late times $u\to +\infty$ the free data behaves as 
\begin{align}
\label{hABufalloff}
        C_{AB}(u, x^A)
        &\overset{u\to+\infty}=
            C_{AB}^{(0),+}(x^A)
            + \frac1u C_{AB}^{(1),+}(x^A)
            + O\Big(\frac{1}{ u^{1+\#}}\Big),
\end{align}
where  $\#$ is ({\it a priori}) any positive number. The $1/\tau$ tail of the Coulombic mode $\overset1h_{\tau \tau}$ on time-like infinity has a counterpart in the $1/u$ behavior of the radiative gravitational field $h_{AB}$ on null infinity which we note violates the property of {\it peeling}.\footnote{The peeling property of asymptotically flat spacetimes translates to asymptotic fall-offs at null infinity of the Weyl scalars $\Psi_k \stackrel{r\to \infty}{=} \Psi_k^0 r^{k-5}+O(r^{k-6})$ where $k=\{1,...,4\}$. Originally conjectured to be a characteristic of the asymptotic radiation emitted by isolated self-gravitating systems \cite{Newman:1961qr,Penrose:1962ij,Friedrich:2017cjg}, there is now growing evidence that peeling is violated in generic gravitational scattering. See \cite{Christodoulou2002, Kehrberger:2021uvf, Geiller:2024ryw, Winicour1985} for details. \label{peeling}} 

The shear mode ${C}^{(0),+}_{AB}$ is sourced by the free matter stress tensor at~$i^+$
\begin{equation}\label{ldmsfms}
    C_{AB}^{(0),+}=-\frac{\kappa^2}{8\pi}\int_{i^+} d^3y \frac{(\partial_A q \cdot \Y) (\partial_B q \cdot \Y)+\frac12\gamma_{AB}}{q\cdot \Y}\overset3T{}^{\rm matt}_{\tau\tau}(y).
\end{equation}
A similar expression can be obtained for the shear mode ${C}^{(0),-}_{AB}$ sourced by the free matter stress tensor at $i^-$.
We now argue that
\begin{equation}
\label{coamatipmcoabatim}
\Delta C^{(0)}_{AB}=C^{(0),+}_{AB} -C^{(0),-}_{AB}
\end{equation}
is precisely the formula for {\it linear displacement memory} sourced by a massive scalar field. 

Typically, the linear displacement memory formula is written in terms of asymptotic momenta of massive particles \cite{Braginsky:1987kwo}. Our expression \eqref{coamatipmcoabatim} has the same functional form except that the asymptotic momenta of the point particles are replaced with a linear momentum flux of the scalar field. To see this, note that linear momentum density of a massive scalar field at $i^{+}$ is given by
\begin{align}\label{lmdaip}
{\cal P}^{\alpha}(y)\, =\, \overset3T{}^{\rm matt}_{\tau\tau}(y) \Y^{\alpha}.
\end{align}
We can use this to express \eqref{ldmsfms} as
\begin{equation}
C_{AB}^{(0),+}=-\frac{\kappa^2}{8\pi}\int_{i^+} d^3y \frac{(\partial_A q \cdot {\cal P}(y)) (\partial_B q \cdot {\cal P}(y))+\frac12\gamma_{AB}\,(\overset3T{}^{\rm matt}_{\tau\tau})^{2}(y)}{q\cdot {\cal P}(y)}.
\end{equation}
This reveals that our formula \eqref{coamatipmcoabatim} is indeed the linear displacement memory formula for a massive field and thus generalises the earlier results for massive point particles.

Gravitational interactions at late times $u\to +\infty$ create a {\it tail} to this memory which is encoded in the $1/u$ term in \eqref{hABufalloff} sourced by the interacting matter stress tensor as well as the graviton stress tensor,
\begin{equation}
\label{Ctail+}
\badat{2}
    C_{AB}^{(1),+}
    &=
        -\frac{\kappa^2}{8\pi}(\p_Aq^\mu)(\p_Bq^\nu)\int_{i^+} d^3y\bigg[
            \frac{
                (q\cdot\Y)\calD^\alpha(\Y_\mu \Y_\nu)
                - (\Y_\mu \Y_\nu+\frac12\eta_{\mu\nu}) \calD^\alpha (q\cdot\Y)
            }{q\cdot \Y}
            \overset{3,\ln}{T}{}^{\rm matt}_{\tau\alpha}
            \\&
            - \left(
                (\Y_\mu \Y_\nu+\frac12\eta_{\mu\nu}) k^{\alpha\beta}
                + (\calD^\alpha\Y_\mu)(\calD^\beta\Y_\nu)
                - \frac12\eta_{\mu\nu}(\calD^\alpha\Y_\sigma)(\calD^\beta\Y^\sigma)
            \right)\overset2T{}^h_{\alpha\beta}
        \bigg]
    .
\eadat
\end{equation}
A similar expression can be obtained for the tail term $C^{(1),-}_{AB}$ sourced by matter and gravitons at $i^-$.
We propose that %
\begin{equation}
 \Delta C^{(1)}_{AB}=C^{(1),+}_{AB}-C^{(1),-}_{AB}
\end{equation}
is the generalization of the {\it tail memory} formula for massive point particles to massive scalar fields. 
We leave a detailed discussion about universal tail terms sourced by matter fields to future work \cite{CLP_memory}.

\section{Superrotation symmetry}
\label{sec:superrotation}

We are now ready to look for a symmetry transformation that is consistent with the logs and tails in the asymptotic behavior of the graviton and matter field. Our starting point is a diffeomorphism vector field $\xi=\xi^\mu \partial_\mu$ which acts as a Lie derivative on the metric on $\scri^+$ and on the real massive scalar field $\varphi$ on $i^+$,
\begin{equation}
\label{deltahvarphi}
    \delta g_{\mu\nu}=\mathcal L_{\xi^{\cal I^\pm}} g_{\mu\nu}%
    , \quad \delta \varphi = \xi^{i^\pm} \cdot \partial \varphi.
\end{equation}
In de Donder gauge the vector field satisfies
\begin{equation}
     \nabla^2 \xi_\mu^{\cal I^\pm}= \nabla^2 \xi_\mu^{i^\pm}=0.
\end{equation}
In the following we will also fix the residual gauge freedom by imposing the radial gauge condition %
\begin{equation}
   X^\mu \nabla_{(\mu} \xi_{\nu)}=0
   .
\end{equation}
We will now identify diffeomorphisms $\xi^{i^\pm}$ and $\xi^{\cal I^\pm}$ on, respectively, time-like and null infinity, consistent with long-range interactions, and show how to smoothly connect them.
The leading and subleading soft graviton theorems can be recast as Ward identities for supertranslations and superrotations whose symmetry parameters on $i^\pm\cup \scri^\pm$ depend only on the angles $x^A$ on the $S^2$. While the associated supertranslation charges are exact, the superrotation charges receive corrections from long-range infrared effects.  We will focus on the future boundary, but analogous expressions hold for the past boundary.

\paragraph{Superrotation on $\scri^+$}
Our ansatz for the large-$r$ limit of the diffeomorphism vector field on $\scri^+$ in de Donder and radial gauge is \cite{Donnay:2020guq}  
\begin{equation}
\label{xiscri+}
\badat{3}
    \xi^{\scri^+}(u,r,x^A) &\stackrel{r\to \infty}=\left[\frac{u}{2} D\cdot Y+O(\tfrac{1}{r})\right]\partial_u - \left[\frac{r}{2}D\cdot Y+\frac{u}{2}\left(\frac{D^2}{2}+2\right)D\cdot Y+O(\tfrac{1}{r})\right]\partial_r \\
    &\qquad + \left[Y^A+\frac{u}{2r}\left((D^2+1)Y^A-D^AD\cdot Y\right)+O(\tfrac{1}{r^2})\right]\partial_A ,
\eadat
\end{equation}
with $D\cdot Y\equiv D_A Y^A$ where $Y^A(\hat x)$ is a vector field on the $S^2$. The Lie derivative along \eqref{xiscri+} on a metric of the form $g_{AB}=r^2\gamma_{AB}+rC_{AB}+\dots$ is given by a superrotation which acts on the sphere metric as
\begin{equation}
\label{superrotationgamma}
\badat{2}
   \delta_Y \gamma_{AB}&=2D_{(A}Y_{B)}-D\cdot Y \gamma_{AB},
\eadat
\end{equation}
and on the gravitational shear as
\begin{equation}
\label{superrotationC}
\badat{2}
   \delta_Y C_{AB}
   &=\left[\mathcal L_Y -\tfrac{1}{2}D\cdot Y (1- u \partial_u) \right]C_{AB}
   \\&\quad
   + u\left[
   D_{(A}(D^2+1)Y_{B)}
   - D_A D_B D\cdot Y
   - \frac12\gamma_{AB}(D^2+4)D\cdot Y
   \right].
\eadat
\end{equation}
Vector fields $Y^A=(Y^z(z),Y^{\bar z}(\bar z))$ that are holomorphic (antiholomorphic) except for isolated points generate Virasoro superrotations \cite{Barnich:2009se,Barnich:2010ojg}, while smooth vector fields $Y^A=Y^A(z,\bar z)$ generate Diff($S^2$) superrotations \cite{Campiglia:2014yka,Campiglia:2015yka,Campiglia:2015kxa}. The enhancement by such superrotations of the BMS group (which is itself an enhancement of the Poincaré group by supertranslations) are referred to as {\it extended} BMS (for Virasoro) and {\it generalized} BMS (for Diff($S^2$)). 

It is worth emphasizing that despite the fact that superrotations act non-trivially on the celestial sphere \eqref{superrotationgamma}, they nevertheless preserve asymptotic flatness as defined by the fall-offs of the Weyl tensor (see footnote \ref{peeling}). Smooth Diff($S^2$) vector fields preserve the peeling property, while meromorphic Virasoro vector fields satisfy a local notion of asymptotic flatness (with topology change at isolated singular points \cite{Strominger:2016wns}).
The novel superrotation charge reported in \cite{Choi:2024ygx}, and shown here to hold to all orders in the coupling $\kappa$, is agnostic about the nature of the vector fields on the $S^{2}$ and thus covers both classes of superrotations.
For an incomplete list of references discussing Virasoro and Diff($S^2$) superrotations, see \cite{Barnich:2009se,Barnich:2010ojg,Kapec:2014opa,Campiglia:2014yka,Kapec:2016jld,Compere:2018ylh,Donnay:2020guq,Geiller:2024amx}.

\paragraph{Superrotation on $i^+$}
Our ansatz for the large-$\tau$ limit of the diffeomorphism vector field on $i^+$ is \cite{Campiglia:2015kxa}
\begin{equation}
\label{xii+}
     \xi^{i^+}(\tau,y) \stackrel{\tau\to \infty}{=}
     \bar Y^\alpha\partial_\alpha 
     + O\left(\tfrac{1}{\tau}\right),
\end{equation}
where $\bar Y^\alpha(y)$ is a vector field on hyperbolic space.
In de Donder gauge it satisfies
\begin{equation}
    \mathcal D_\alpha \bar Y^\alpha=0.
\end{equation}
It acts on the real matter field as a superrotation
\begin{equation}
    \delta \varphi =\bar Y^\alpha \partial_\alpha \varphi.
\end{equation}

\paragraph{Superrotation on $i^+\cup \scri^+$}
The last step is to identify a superrotation that smoothly interpolates across the union of the future time-like and null boundaries $i^+\cup \scri^+$. To this end, note that  the vector field $\bar Y^\alpha$ on $i^+$ is given in terms of the vector field $Y^A$ on $\calI^+$ \cite{Campiglia:2015lxa},
\begin{equation}
    \bar Y^\alpha(y)
    =
        \int_{S^2} d^2x \,
        G^\alpha_A(y;\hat x) Y^A(\hat x)
        ,
    \label{YbarY}
\end{equation}
where $ G^\alpha_A(y;\hat x)$ is the bulk-to-boundary vector Green's function (see Appendix~\ref{app:Green})
which satisfies
\begin{equation}
\label{D2-2Green}
   (\calD^2-2)G^\alpha_A(\rho,\hat x;\hat x') =0.
\end{equation}
In the large-$\rho$ limit it behaves as 
\begin{equation}
    \lim_{\rho\to\infty} 
    G^A_B(\rho,\hat x;\hat{x}') = \delta^A_B\delta^{2}(\hat x-\hat x').
\end{equation}
At the boundary of $i^+$, using \eqref{taurho}, we then find
\begin{equation}
    \label{Yi+}
    \xi^{i^+}(\tau,\rho,\hat x)
    \overset{\rho\to \infty,\tau\to \infty}{=}  Y^A\p_A+\dots\,.
\end{equation}
Choosing the same vector field $Y^A(\hat x)$ at the boundary of $i^+$ we have at leading order
\begin{equation}
    \bar Y^A(\rho,\hat x)
    \overset{\rho\to \infty}{=}  Y^A(\hat x).
\end{equation}
Thus we have identified a superrotation vector field that smoothly extends across the future boundary $i^+\cup \scri^+$. Such a smooth extension is indeed expected as we consider the action of asymptotic symmetry generators on an asymptotically flat metric with a smooth gluing of the (neighbourhood) of $i^{\pm}$ with retarded Bondi co-ordinates as $u\, \rightarrow\, \infty$.

\section{Symplectic structure and charges in gravity%
}
\label{sec:SymplecticStructure}

For the asymptotics of the matter and radiative fields discussed in section \ref{sec:AsymptoticsGravity} which account for the long-range nature of the interactions, we can now compute the symplectic structure
\begin{equation}
    \Omega_{i^+\cup \scri^+}(\delta,\delta')=\Omega^{\rm mat}_{i^+}(\delta,\delta')+\Omega^{\rm rad}_{\scri^+}(\delta,\delta').
\end{equation}
The more relaxed fall-offs of the fields in the form of late-time logs and tails will lead to infrared divergences in the symplectic structure which we will regularize.
After taking one of the variations to correspond to the superrotation identified in section~\ref{sec:superrotation}, we will be able to extract (regularized) charges from the symplectic structure
\begin{equation}
\Omega_{i^+\cup \cal I^+}(\delta,\delta_Y)=\delta Q_{+}[Y].
\end{equation}
Again we focus on the future, but a similar analysis can be repeated in the past.
Our final goal is to extract conservation laws that account for the long-range nature of the interactions and match onto the logarithmic soft graviton theorem.
Moreover, we will show that our new conservation laws are exact to all orders in the gravitational coupling $\kappa$.

\subsection{Matter symplectic structure %
and hard charge %
}
\label{sec:MatterOmega}
In the following we derive the (asymptotic) symplectic structure on the space of solutions of a massive real scalar field minimal coupled to Einstein gravity\footnote{Note that we have omitted perturbative gravity corrections to the symplectic structure, $\Omega_\tau^{\rm corr}$, since these vanish as we approach the time-like boundaries $i^\pm$, with fall-off conditions on the fields (and their variations) given in \eqref{varphiati+}.
To see this, consider the leading gravitational correction to the symplectic structure of the asymptotically free field 
\begin{align}
    \lim_{\tau\to \infty} \Omega^{\rm corr}_{\tau}(\delta, \delta^{\prime})=\lim_{\tau \to \infty} \tau^{3} \delta \int_{{\cal H}_{\tau}} d^3y\,h^{\tau\mu}\partial_{\mu}\varphi \, \delta'\varphi -(\delta \leftrightarrow\delta').
\end{align}
As $\tau \to \infty$ we have $h^{\tau\mu}\partial_{\mu}\varphi \sim O(\tau^{-5/2}\ln\tau)$ and as a result 
\begin{align}
 \lim_{\tau\to \infty} \Omega^{\rm corr}_{\tau}(\delta, \delta^{\prime})=0.
\end{align}
This argument depends on the variation of the matter field falling off no slower than $\delta\varphi\sim \ln\tau/\tau^{3/2}$ which is true for the variation generated by superrotation symmetries considered in this work. In contrast, higher spin symmetries such as those associated which the sub-subleading soft graviton theorem will in fact lead to $\delta\varphi = O(\tau^{-1/2}{\ln\tau})$ and hence for such symmetries $\Omega^{\rm corr}$ will contribute to the matter symplectic structure.}
\begin{equation}
    \Omega_{\tau}=\int_{\Sigma_{\tau}} dS_\mu \omega^\mu,\quad \omega^\mu=\delta \varphi {\nabla}{}^\mu \delta'\varphi-(\delta \leftrightarrow\delta').
\end{equation}
On a $\tau$ = constant hyperbolic slice $\mathcal H_\tau$ the symplectic structure of a free massive real scalar field is given by
\begin{align}
\label{GravityOmegataufree}
    \Omega^{\rm mat,free}_\tau
    &=
        \int_{\cal H_\tau} d^3y\,\tau^3
        \omega^\tau_{\rm free}
    ,\qquad
    \omega^\tau_{\rm free}
    =
        - \delta\varphi_\text{free} \partial_\tau\delta'\varphi_\text{free}
        -(\delta \leftrightarrow \delta').
\end{align}
The free-field symplectic structure defined in terms of the vector space of free-data at time-like infinity is the well-known Fock space symplectic structure
\begin{equation}
\label{Omegai+free}
         \Omega^{\rm mat,free}_{i^+}= \lim_{\mathcal H_\tau \to i^+}\Omega^{\rm mat, free}_{\tau}= %
2im\int_{i^+} d^3y
        \left[
            \delta b_0^*  \delta' b_0
           -(\delta \leftrightarrow \delta')
        \right].
\end{equation}
In four spacetime dimensions long-range interactions dress the asymptotic free fields by `Coulombic tails' which will give rise to infrared divergences that we have to regularize. %

Our starting point is thus the symplectic structure of a massive real scalar field %
\begin{align}
\label{GravityOmegatau}
    \Omega^{\rm mat}_\tau
    &=
        \int_{\cal H_\tau} d^3y\,\tau^3
        \omega^\tau
    ,\qquad
    \omega^\tau(\delta,\delta')
    =
        -\delta\varphi \partial_\tau\delta'\varphi
        -(\delta \leftrightarrow \delta'),
\end{align}
whose $\tau \to \infty$ asymptotics accounts for long-range interactions.
For the scalar field expansion \eqref{varphiati+}  with \eqref{Gravity_expansion} accounting for these infrared effects, we find
\begin{equation}
\label{Gravityomegatau}
    \omega^{\tau}(\delta,\delta')
    =
        2im
        \Big\{
            \ln\tau\Big[
                \delta \overset\ln b{}_0^* \delta' b_0
                + \delta b_0^* \delta' \overset\ln b{}_0
            \Big]+ \Big[\delta b_0^* \delta' b_0\Big]
            +\dots\Big\}
        -(\delta\leftrightarrow \delta')
    .
\end{equation}
Note that this expression is exact in the coupling $\kappa$.

We evaluate the symplectic structure on $i^+$ by taking the late-time limit $\tau \to \infty$,
\begin{equation}
    \Omega_{i^+}^{\rm mat}(\delta,\delta')=\lim_{\mathcal H_\tau \to i^+} \Omega_\tau^{\rm mat}(\delta,\delta').
\end{equation}
To regulate the $\ln \tau$ infrared divergence in \eqref{Gravityomegatau}
we introduce a late-time cutoff $\Lambda^{-1}$. This will allow us to extract observables that remain finite when we eventually remove the cutoff $\Lambda\to 0$ (corresponding to $\tau\to \infty$).
When one variation is taken to be a superrotation with vector field \eqref{xii+}, we can extract from
\begin{equation}
\label{deltaQH}
\Omega^{\rm mat,reg}_{i^+}(\delta,\delta_{\bar Y})=\delta Q^\Lambda_{H,+}[\bar Y]
\end{equation}
the {\it hard} Noether charge 
\begin{equation}
\badat2
\label{GravityOmegamati+renfinal}
     Q^\Lambda_{H,+}[\bar Y]
     &=
        im
        \int_{i^+} d^3y\, \bar Y^\alpha\left\{
            2\ln\Lambda^{-1}
            \Big[
                b_0^* \p_\alpha \overset\ln b{}_0
                - b_0 \p_\alpha\overset\ln b{}_0^*
            \Big]+
        \Big[
             b_0^* \p_\alpha b_0
             - b_0 \p_\alpha b_0^*
             + O(\kappa^2)
        \Big]
        \right\},
\eadat
\end{equation}
where we used integration by parts and the property $\calD_\alpha \bar Y^\alpha=0$.
Expressing this as
\begin{equation}\label{GravityQH+}
   Q^\Lambda_{H,+}[\bar Y]= \ln \Lambda^{-1}Q^{(\ln)}_{H,+} [\bar Y]+Q^{(0)}_{H,+}[\bar Y],
\end{equation}
we identify two contributions to the {\it hard} matter charge on the future time-like boundary.
The `hard log charge' is given by
\begin{equation}\label{GravityQHln+}
\badat2
    Q^{(\ln)}_{H,+}[\bar Y]
    &=
    -\frac{\kappa m^3}{4(2\pi)^3} \int_{i^+} d^3y\, \bar Y^\alpha (\p_\alpha\overset1h_{\tau\tau})b^*b
    ,
\eadat
\end{equation}
which we stress is {\it exact in the gravitational coupling $\kappa$}, 
while the coefficient of the cutoff-independent term is given by 
\begin{equation}\label{GravityQH0+}
\badat2
    Q^{(0)}_{H,+}[\bar Y]
    &=
    \frac{im^2}{4(2\pi)^3}\int_{i^+} d^3y\, \bar Y^\alpha
    (b^*\p_\alpha b-b\p_\alpha b^*)+ O(\kappa^2)
    .
\eadat
\end{equation}
Here we have used $b_0=\sqrt{\frac{m}{4(2\pi)^3}}b$ to write the final expressions in terms of modes that satisfy the standard Poisson bracket $i\{b(\vec p),b(\vec p')\}_{\rm PB} =(2\pi)^3(2E_p)\delta^3(\vec p-\vec p')$, or in hyperbolic coordinates $\vec p=m\rho\hat x$,
\begin{equation}\label{canonicalPB}
    i\{b(y),b(y')\}_{\rm PB} =(2\pi)^3\frac{2}{m^2}\delta^3(y-y')
    .
\end{equation}

Analogous expressions are obtained at the past time-like boundary $i^-$.

\subsection{Radiative symplectic structure %
and soft charge %
}
\label{sec:RatiativeOmega}
The radiative symplectic structure is given by
\begin{equation}
\Omega(\delta,\delta') =\int_\Sigma dS_\mu \omega^\mu(\delta,\delta') , 
\end{equation}
with pre-symplectic form  
\begin{equation}
\badat{2}
    \omega^\mu(\delta, \delta')&=
		\frac{1}{\kappa^2}\Big[
        -g^{\sigma\lambda}\delta g_{\sigma\lambda}\nabla_\nu \delta'g^{\mu\nu}
		- g^{\sigma\lambda}\delta g_{\sigma\lambda}\nabla^\mu(g^{\nu\kappa}\delta'g_{\nu\kappa})
		- \delta g^{\mu\nu} \nabla_\nu (g^{\sigma\lambda}\delta'g_{\sigma\lambda})
    \\&\qquad\qquad
		+ 2\delta g_{\nu\sigma}\nabla^\nu \delta'g^{\mu\sigma}
		- \delta g_{\nu\sigma}\nabla^\mu \delta'g^{\nu\sigma}
		- (\delta\leftrightarrow \delta')
        \Big]
    .
\eadat
\end{equation}
On a Cauchy slice $\Sigma_t$ of constant $t$ with $dS_\mu= d^3x \,n_\mu$ and normal $n_\mu=\partial_\mu t$ the symplectic structure  is $\Omega_{\Sigma_t}=\int_{\Sigma_t} d^3x \,\omega^t $. 
We can evaluate this on future null infinity by going to retarded Bondi coordinates $u=t-r$ %
and pushing $r\to\infty$ while holding $u$ fixed,
\begin{equation}
    \Omega_{\scri^+}=\lim_{\Sigma_t\to \scri^+}\Omega_{\Sigma_t}%
    =-\int_{\scri^+} du d^2 \hat x\, r^2 \omega_u
    .
\end{equation}

For a graviton field with `standard' large-$r$ fall-offs \eqref{standardhlarger} and \eqref{hABlarger}, and for variations that respect these fall-offs, the radiative symplectic structure is simply
\begin{equation}
    \Omega^{\rm rad,AS}_t=\int_{\Sigma_t} d^3x \,\omega^t,\qquad \omega^t=  -\frac{1}{\kappa^2} \delta C_{AB} \partial_t\delta' C^{AB} - (\delta \leftrightarrow \delta').
\end{equation}
Upon pushing the Cauchy slice $\Sigma_t$ to future null infinity this
becomes \cite{Frolov:1977bp,Ashtekar:1981bq}\begin{equation}
\label{GravityOmegaradASscri+}
     \Omega^{\rm rad,AS}_{\scri^+}=\lim_{\Sigma_t\to \scri^+} \Omega^{\rm rad,AS}_t= -\frac{1}{\kappa^2} \int_{\scri^+}  du d^2\hat x \left[ \delta C_{AB} \partial_u\delta' C^{AB} - (\delta \leftrightarrow \delta')\right].
\end{equation}
This is the famous Ashtekar-Streubel (AS) radiative symplectic structure.
It is finite provided that the shear at early and late times falls off as
\begin{equation}
    C_{AB}(u,\hat x)\stackrel{u\to \pm \infty}{=} C^{(0),\pm}_{AB}(\hat x)+O\Big(\frac{1}{|u|^\#}\Big),
\end{equation} 
where $\#$ is any positive number. %
While this includes the asymptotic behavior~\eqref{hABufalloff} involving $1/u$ tails, the Ashtekar-Streubel symplectic structure cannot account for superrotations which change the metric on the celestial sphere as  \eqref{superrotationgamma}.

The superrotation action on the angular components of the metric at leading order turns out to render the radiative symplectic structure linearly divergent in $r$,
\begin{equation}
	\label{GravityOmegascri+}
    \Omega_{\scri^+}^{\rm rad}=-\int_{\scri^+} du d^2 \hat x\, r^2 \omega_u,\qquad r^2\omega_u=r \,\overset{\rm div}{\omega_u}+\overset{\rm fin}{\omega_u}
    .
\end{equation}
This $O(r)$ divergence is well-known in the literature \cite{Compere:2018ylh, Donnay:2020guq,Campiglia:2015yka} and can be removed via an appropriate renormalization of $\omega$. Expressing the pre-symplectic form as $\omega(\delta,\delta')=\delta \Theta(\delta')-(\delta\leftrightarrow \delta')$ exposes the freedom to perform a shift in the definition of the pre-symplectic potential~$\Theta \mapsto \Theta+d\theta$ where $\theta$ is a spacetime co-dimension 2 form. After exploiting this freedom to remove the linear in $r$ divergence, our starting point is the radiative symplectic structure \cite{Donnay:2020guq}
\begin{equation}\label{GravityOmegaradscri+}
\Omega^{\rm rad}_{\scri^+}
    =
    -\frac{1}{\kappa^2}\int_{\cal I^+} dud^2\hat x
    \left[
    \delta C^{AB}\p_u \delta' C_{AB}
    + \frac12 \delta \gamma^{AB}D^2\delta' C_{AB} - (\delta\leftrightarrow \delta')\right].
\end{equation}

If one of the variations is a superrotation \eqref{xiscri+} the radiative symplectic structure~\eqref{GravityOmegaradscri+} for a gravitational field with the long-range tails~\eqref{hABufalloff} %
turns out to be a total variation in field space,
\begin{equation}\label{GravitydeltaQraddrag}
	\Omega^{\rm rad}_{\scri^+}(\delta,\delta_Y)
	=
	\delta Q^{\rm rad}_{S,+}[ Y]
	,
\end{equation}
which allows us to extract the charge
\begin{equation}
\badat2 \label{GravityQrad}
    Q^{\rm rad}_{S,+}[Y]
    &=
    \frac{1}{\kappa^2}
    \int_{\cal I^+} dud^2\hat x
    \Big[
        (1-u\partial_u)C^{AB}
        \left(
            2D_AY_B
            + D_AD_BD\cdot Y
        \right)+u N^{AB} D^2D_AY_B
    \Big].
    \eadat
\end{equation}
Acting with a superrotation results in the charge \eqref{GravityQrad} being linearly divergent in $u$,
which we can remove by adding a suitable boundary term,
\begin{equation}
\badat{2}
    Q^{\rhd}_{S,+}[Y]
    &=
    \frac{1}{\kappa^2}
    \int_{\cal I^+} dud^2\hat x\,
    \p_u\left[
        -u C^{AB}
        \left(
            2D_AY_B
            + D_AD_BD\cdot Y
        \right)
    \right]
    ,
    \eadat
\end{equation}
which is a corner term on $\scri^+$.\footnote{
Contributions to the symplectic structure which are localised on co-dimension two boundaries can be interpreted as the inherent ambiguity in the definition of symplectic structure.
Hence the removal of the linearly divergent term $Q^{\rhd}_{S,+}[Y]$ is equivalent to fixing such a corner term ambiguity.}
This yields
\begin{equation}
\badat2
\label{GravityQradreg}
     Q^{\rm rad+\rhd}_{S,+}[Y]&=Q^{\rm rad}_{S,+}[Y]+ Q^{\rhd}_{S,+}[Y]\\
    &=
    - \frac{1}{\kappa^2}
    \int_{\cal I^+} dud^2\hat x\,
        uN^{AB}
        \left[
            2D_AD_BD\cdot Y
            - (D^2-4)D_AY_B
        \right]
    .
\eadat
\end{equation}
The $u$-integral in \eqref{GravityQradreg} along the null direction is logarithmically divergent but this is due to the long-range nature of the interactions.
We regulate this divergence at early and at late times $u \to \pm \infty$ via an infrared cutoff $\Lambda^{-1}$:
We pick a finite but large $u_0$, and split the $u$-integral into three segments $(-\Lambda^{-1},-u_0)\cup (-u_0,u_0)\cup(u_0,\Lambda^{-1})$, so that the contribution from the middle segment is finite.\footnote{This is true as long as $C_{AB}(u,\hat{x})$ is analytic in $(-\infty, \infty)$ and this is indeed true for smooth classical asymptotic data at ${\cal I}^{\pm}$. If the asymptotic data is distributional (for example the data associated to shockwave geometries) then the charge in $(-\Lambda^{-1},\, \Lambda^{-1})$ will have a divergence, but we do not consider such a scenario in this paper.}
We may approximate the integrand of the upper and lower segments using \eqref{hABufalloff}, with an error of order $O(u_0^{-1})$ which is finite since $u_0$ is finite (and large).
This implies that the integrals are logarithmically divergent in $\Lambda^{-1}$,
\begin{equation}
\badat2
    \int_{-\Lambda^{-1}}^{+\Lambda^{-1}} du \, u\p_u C_{AB}
    &=
        - \int_{+u_0}^{+\Lambda^{-1}} du \Big(\frac1u C^{(1)}_{AB}\Big)
        - \int_{-\Lambda^{-1}}^{-u_0} du \Big(\frac1u C^{(1)}_{AB}\Big)
        + {\rm (finite)}
    \\ &=
        - \ln \Lambda^{-1}\Big(C^{(1),+}_{AB}-C^{(1),-}_{AB}\Big)
        + O(\Lambda^0)
    .
\eadat
\end{equation}
One may readily check from \eqref{hABufalloff} that the expression in parentheses correspond to the integral of $-\p_u (u^2\p_u C_{AB})$,
\begin{equation}
    \int^{+\Lambda^{-1}}_{-\Lambda^{-1}}du\, u\p_u C_{AB}
    =
        \ln\Lambda^{-1}\int_{-\infty}^\infty du\, \p_u(u^2\p_uC_{AB})
        + O(\Lambda^0)
    .
\end{equation}

Long-range interactions now add a drag: due to the spacetime curvature caused by the matter the soft graviton experiences a {\it gravitational drag} at late times.
This effect can be taken into account by solving Einstein's equations at $\cal I^+$.
The gravitational drag on photons takes the form of a phase that is logarithmically divergent in $r$ \cite{AtulBhatkar:2019vcb}. By the equivalence principle gravitons undergo the same effect, which results in the shift 
\begin{equation}
    \lim_{u\, \rightarrow\,\Lambda^{-1}}\, C_{AB}(u,\hat x)
    \quad\to\quad
    \lim_{u\, \rightarrow\, \Lambda^{-1}}\, \left[\, C_{AB}(u,\hat x)
    - \frac\kappa 2 \ln \Lambda^{-1}\,\overset0h_{rr}(\hat x) \p_u C_{AB}(u,\hat x)\, \right]
    .
\end{equation}
The shift in $C_{AB}$ amounts to an additional contribution to the charge
\begin{equation}
\badat2
\label{GravityQdrag}
     Q^{\rm drag,reg}_{S,+}[Y]=
     \frac{1}{2\kappa}\ln \Lambda^{-1}
     \int_{\cal I^+} dud^2\hat x\,
        \overset0h_{rr}\partial_u C^{AB}
        \left[
            2D_AD_BD\cdot Y
            - (D^2-4)D_AY_B
        \right]
    ,
\eadat
\end{equation}
where we have integrated by parts in $u$ and used the fact that the resulting boundary term vanishes since $\p_u C_{AB}$ falls off as $O(u^{-2})$, as can be seen from \eqref{hABufalloff}.\footnote{The integration range ${\cal I}^{+}$ in \eqref{GravityQdrag}  should be understood as being regularised by the cutoff $\Lambda^{-1}$. However as the integral is not divergent as $\Lambda^{-1}\, \rightarrow\, \infty$, we simply replace $(-\Lambda^{-1}, \Lambda^{-1})$ with ${\cal I}^{+}$.}

After this cutoff regularization we can then define the {\it soft} charge on the future null boundary 
\begin{equation}
\label{GravityOmegarenscri+}
    \badat{2}
	Q^\Lambda_{S,+}[Y]&= Q^{\rm rad+\rhd,reg}_{S,+}[Y]
	+ Q^{\rm drag,reg}_{S,+}[ Y]
\\  
    &=\ln \Lambda^{-1} Q_{S,+}^{(\ln)}[Y]+Q_{S,+}^{(0)}[Y],
    \eadat
\end{equation}
where we identify two distinct contributions.
The `soft log charge' is 
\begin{equation}
\badat2
\label{GravityQSln+}
    Q_{S,+}^{(\ln)}[Y]
     &=
     - \frac{1}{\kappa^2}
     \int_{\cal I^+} dud^2\hat x\,\left[
        \p_u (u^2\p_u C^{AB})-\frac{\kappa}{2}\overset0h_{rr}\partial_u C^{AB}\right]\\
        &\qquad \qquad \qquad\qquad\times
        \left[
            2D_AD_BD\cdot Y
            - (D^2-4)D_AY_B
        \right]
    ,
\eadat
\end{equation}
with $C_{AB}$ given in \eqref{hABufalloff}. The coefficient of the cutoff-independent term is 
\begin{equation}
\label{GravityQS0+}
    Q_{S,+}^{(0)}[Y]
    =
     -\frac{1}{\kappa^2}
     \int_{\cal I^+} dud^2\hat x
       \, u \p_u C^{AB}
        \left[
            2D_AD_BD\cdot Y
            - (D^2-4)D_AY_B
        \right],
\end{equation}
where, in slight abuse of notation, $C_{AB}$ now denotes the early/late time fall-offs \eqref{hABufalloff} without the Coulombic tail. 
Analogous expressions are obtained at the past null boundary~$\scri^-$.

\section{From charge conservation laws to soft graviton theorems}
\label{sec:Gravitycharges}

We are now finally in a position to establish the connection between the conservation laws for superrotation symmetry and the logarithmic soft graviton theorem. As a byproduct of our analysis, we will see that superrotation symmetry continues to imply also the subleading tree-level soft graviton theorem. For comparison we will also review the relation between supertranslation symmetry and the leading soft graviton theorem.  

It will be convenient to express the hard charges in terms of the energy-momentum tensor. 
The charge that generates a diffeomorphism $\xi$ on $i^+$ can be written as \cite{Campiglia:2015kxa}
\begin{equation}
    Q_{H,+}[\xi]
    =
        - \int_{i^+} dS_\mu T^\mu{}_\nu \xi^\nu.
\end{equation}
For supertranslations $\xi=\bar f\p_\tau$ and superrotations $\xi = \bar Y^\alpha\p_\alpha$, the expressions for their charges at $i^+$ are given, respectively, by
\begin{align}
    Q_{H,+}[\bar f]
    &=
        \lim_{\tau\to\infty} \int_{\cal H_\tau} d^3y\,\tau^3 T_{\tau\tau} \bar f
    ,\\
    Q_{H,+}[\bar Y]
    &=
        \lim_{\tau\to\infty} \int_{\cal H_\tau} d^3y\,\tau^3 T_{\tau\alpha}\bar Y^\alpha
    ,
    \label{QY}
\end{align}
and similar expressions at $i^-$.

\subsection{Leading soft theorem}
We start by briefly reviewing the symmetry interpretation of the leading soft graviton theorem \cite{He:2014laa,Campiglia:2015kxa}
\begin{equation}
\lim_{\omega\to 0}\omega {\cal M}_{N+1}=
    S_{-1} {\cal M}_N.
\end{equation}
For a supertranslation $\xi^{\scri^+}=f(\hat x)\partial_u$ at null infinity such that the shear transforms as $ \delta C_{AB}(u,\hat x)=-(2D_A D_B-\gamma_{AB} D^2) f(\hat x)$ we get the expression for the soft charge
\begin{equation}
\label{gravQSleading}
     Q^{(-1)}_{S,\pm}[f] =-\frac{1}{\kappa^2}\int_{S^2}   d^2\hat x \,(2D^A D^B-\gamma^{AB} D^2) f(\hat x) \, \mathcal J^{(-1)}_{AB,\pm}
\end{equation}
in terms of the leading soft graviton operator on $\scri^+$,  
\begin{equation}
    \label{J-1AB}
   \mathcal J^{(-1)}_{AB,+}\equiv \int_{-\infty}^{+\infty} du \,\partial_u C_{AB}(u,\hat x)\quad \text{with} \quad C_{AB}\stackrel{u\to \pm\infty}=C_{AB}^{(0),\pm}+O(\frac{1}{|u|^{\#}}),
\end{equation}
where $\#$ is any positive number; a similar expression is obtained on $\scri^-$.
For $A,B=z,\bar z$ we recover the expressions
\begin{equation}
\label{GravityQS-1+zzbar}
    Q_{S,\pm}^{(-1)}[Y]
    =
     -\frac{2}{\kappa^2}
     \int_{S^2} d^2\hat x
        \left[D_z^2 f \mathcal  J^{(-1),zz}_{+} +D_{\bar z}^2f \mathcal J^{(-1),\bar z \bar z}_{+}\right]
        ,
\end{equation}
familiar from the literature \cite{Strominger:2013jfa,He:2014laa}.
For a supertranslation $\xi^{i^+}=\bar f(y)\p_\tau$ at time-like infinity such that $\delta \varphi(\tau,y)=\bar f(y)\p_\tau \varphi$, we get the expression for the hard charge \cite{Campiglia:2015kxa}
\begin{equation}
    \label{gravQHleading}
    Q^{(-1)}_{H,\pm}[\bar{f}] =\int_{i^\pm} d^3 y \,\bar{ f}(y) \,\overset3T{}^{\rm matt}_{\tau\tau}(y)
\end{equation}
in terms of the free matter stress tensor \eqref{Tmatt}. Using $b_0=\sqrt{\frac{m}{4(2\pi)^3}}b$ we have
\begin{equation}\label{T3tt}
    \overset3T{}_{\tau \tau}^{\rm matt}=%
    \frac{m^3}{2(2\pi)^3}b^*b
    ,
\end{equation}
where $b$ and $b^*$ are the scalar modes that satisfy the canonical bracket \eqref{canonicalPB}.

Both soft and hard charges are exact in the gravitational coupling $\kappa$. 
The superscripts $(-1)$ anticipate the connection of these soft and hard charges with the leading soft graviton theorem which scales as $\omega^{-1}$ in the soft expansion. Indeed, upon antipodal identification of the symmetry parameters $\bar{f}(y)$ and $f(\hat x)$ between $i^+\cup \scri^+$ and $i^-\cup \scri^-$, we obtain charges
\begin{equation}
 Q^{(-1)}_\pm=  Q^{(-1)}_{S,\pm}+ Q^{(-1)}_{H,\pm}
\end{equation}
on the future ($+$) and past ($-$) boundary that satisfy
\begin{equation}
     Q^{(-1)}_+=Q^{(-1)}_-.
\end{equation}
Upon quantization this charge conservation law for supertranslation symmetry corresponds to the leading soft graviton theorem~\cite{Campiglia:2015kxa}.\footnote{See ~\cite{He:2014laa} for the relation between supertranslations and the soft graviton theorem for massless particles.} 

\subsection{Subleading tree-level soft theorem}
\label{sec:Gravitysubtreecharges}
In \cite{Kapec:2014opa,Campiglia:2015kxa} it was shown that the subleading tree-level soft graviton theorem
\begin{equation}
\lim_{\omega\to 0}(1+\omega \partial_\omega) {\cal M}_{N+1}=
    S_{0} {\cal M}_N
\end{equation}
can also be understood from a conservation law of charges
\begin{equation}
     Q^{(0)}_+=Q^{(0)}_-
\end{equation}
associated to superrotations \eqref{xiscri+} and \eqref{xii+}. The hard and soft contributions to the future~($+$) and past~($-$) charges,
\begin{equation}
 Q^{(0)}_\pm=  Q^{(0)}_{S,\pm}+ Q^{(0)}_{H,\pm},
\end{equation}
are precisely \eqref{GravityQH0+} and \eqref{GravityQS0+}. So, as a byproduct of our analysis which took into account the long-range nature of gravitational interactions, we rediscover the result of \cite{Campiglia:2015kxa}.\footnote{See~\cite{Kapec:2014opa} for the relation between superrotations and the subleading soft graviton theorem for massless particles.}

The soft charge is given by 
\begin{equation}
    Q_{S,\pm}^{(0)}[Y]=-\frac{1}{\kappa^2}\int_{S^2}  d^2\hat x \left[2D^A D^B D\cdot Y -(D^2-4)D^A Y^B\right]  \,\mathcal J^{(0)}_{AB,\pm}
\end{equation}
in terms of the subleading soft graviton operator on $\scri^+$ 
\begin{equation}
    \mathcal  J^{(0)}_{AB,+}\equiv  \int_{-\infty}^{+\infty} du\, u\partial_u  C_{AB} \quad \text{with} \quad C_{AB}\stackrel{u\to \pm\infty}{=}C^{(0),\pm}_{AB}+O(\frac{1}{|u|^{\#}}),
\end{equation}
where $\#$ is any positive number $>1$; a similar expression is obtained on $\scri^-$. For $A,B=z,\bar z$ we recover the form
\begin{equation}
    Q_{S,\pm}^{(0)}[Y]
    =
     -\frac{2}{\kappa^2}
     \int_{S^2} d^2\hat x
        \left[D_z^3Y^z  \mathcal  J^{(0),zz}_{+} +D_{\bar z}^3Y^{\bar z} \mathcal J^{(0),\bar z \bar z}_{+}\right]
        ,
\end{equation}
familiar from the literature \cite{Kapec:2014opa,Campiglia:2015kxa}.
The hard charge can be written as
\begin{align}
	Q_{H,\pm}^{(0)}[\bar Y]
	&=
   \int_{i^\pm} d^3y\, \bar Y^\alpha(y)
		\overset3T{}_{\tau\alpha}(y),
\end{align}
where the energy-momentum tensor obtains contributions from matter and gravitons,
\begin{equation}
    \overset3T_{\tau\alpha} = \overset3T{}^{\rm matt}_{\tau\alpha} + \overset3T{}^h_{\tau\alpha}
    .
\end{equation}
The graviton contribution is at least of order $O(\kappa^2)$, while the matter contribution \eqref{Tmatt} has a free part of order $\kappa^0$ and no interacting part of order $O(\kappa^2)$ so that
\begin{equation}\label{T3tamatt}
    \overset3T{}_{\tau\alpha}^{\rm matt}=%
    \frac{im^2}{4(2\pi)^3} (b^*\p_\alpha b-b\p_\alpha b^*)
\end{equation}
is actually exact.
We have again used $b_0=\sqrt{\frac{m}{4(2\pi)^3}}b$ to write the matter stress tensor  \eqref{Tmatt} in terms the modes $b$, $b^*$ that satisfy the canonical bracket \eqref{canonicalPB}.

\subsection{Logarithmic soft theorem}
\label{sec:Gravitylogcharges}

Our goal in this work is to identify a conservation law of charges derived from a first principles covariant phase space approach which is equivalent to the classical logarithmic soft graviton theorem 
\begin{equation}
  \lim_{\omega \to 0}\partial_\omega \left(\omega^2 \partial_\omega {\cal M}_{N+1}\right)=
    S^{(\ln \omega)}_{0,{\rm classical}} {\cal M}_N.
\end{equation}
We expect the latter to be associated with the same diffeomorphism \eqref{xiscri+} and \eqref{xii+} as the tree-level subleading soft graviton theorem corresponding to superrotations but with an infrared-corrected Noether charge given by our soft and hard logarithmic charges \eqref{GravityQSln+} and \eqref{GravityQHln+}. The soft log charge is given by 
\begin{equation}
\badat2
\label{grQSln}
    Q_{S,+}^{(\ln)}[Y]
    &=
    -\frac{1}{\kappa^2}\int_{S^2} d^2\hat x \left[2D^A D^B D\cdot Y-(D^2-4)D^A Y^B \right] \,    \mathcal J^{(\ln)}_{AB,+}
    \\&\quad
    +\frac{1}{2\kappa}
    \int_{S^2} d^2\hat x\,
        \overset0h_{rr}
        \left[
            2D^AD^BD\cdot Y
            - (D^2-4)D^AY^B
        \right]
        {\cal J}^{(-1)}_{AB,+}
\eadat
\end{equation}
in terms of the leading soft graviton operator \eqref{J-1AB} and the `log soft graviton operator' 
\begin{equation}
    \mathcal J^{(\ln)}_{AB,+}\equiv \int_{-\infty}^{+\infty} du\,\partial_u(u^2 \partial_u  C_{AB})\quad \text{with} \quad C_{AB}\stackrel{u\to \pm\infty}{=}C^{(0),\pm}_{AB}+\frac{1}{|u|}C^{(1),\pm}_{AB}+O(\frac{1}{|u|^{1+\#}})
\end{equation}
and $\#$ any positive number. For $A,B=z,\bar z$ we find\footnote{Note that our expression for the soft charge agrees with the BRST cocycle analysis of \cite{Baulieu:2024oql}. We thank Tom Wetzstein for pointing this out to us.}
\begin{equation}
\badat2
    Q_{S,\pm}^{(\rm ln)}[Y]
    &=
     -\frac{2}{\kappa^2}
     \int_{S^2} d^2\hat x
        \left[D_z^3Y^z  \mathcal  J^{(\ln),zz}_{+} +D_{\bar z}^3Y^{\bar z} \mathcal J^{(\ln),\bar z \bar z}_{+}\right]
    \\&\quad
    +\frac{1}{\kappa}
    \int_{S^2} d^2\hat x\,
        \overset0h_{rr}
        \left[D_z^3Y^z  \mathcal  J^{(-1),zz}_{+} +D_{\bar z}^3Y^{\bar z} \mathcal J^{(-1),\bar z \bar z}_{+}\right]
    ,
\eadat
\end{equation}
where $d^2\hat x= \gamma_{z\zb} dzd\zb$.
The hard log charge can be written as
\begin{align}\label{grQHln}
	Q_{H,+}^{(\ln)}[\bar Y]
	&=
		 \int_{i^+} d^3y\, \bar Y^\alpha
		\overset{3,\ln}T{}^{\rm matt}_{\tau\alpha},
\end{align}
where the expression for the interacting stress energy tensor \eqref{Tmatt}
using $b_0=\sqrt{\frac{m}{4(2\pi)^3}}b$ is given by
\begin{equation}\label{T3lntamatt}
    \overset{3,\ln}T{}_{\tau \alpha}^{\rm matt}=%
   - \frac{\kappa m^3}{4(2\pi)^3}(\p_\alpha\overset1h_{\tau\tau}) b^* b
    ,
\end{equation}
with the $b$ and $b^*$ modes satisfying the canonical Poisson bracket \eqref{canonicalPB}.
{\it Both soft and hard logarithmic charges are exact in the gravitational coupling $\kappa$.}

One obtains soft and hard charges, $Q^{(\ln)}_{S,-}$ and $Q^{(\ln)}_{H,-}$, on the past null and time-like infinities, respectively, by a similar analysis.
Upon antipodal identification of the symmetry parameters $\bar{ Y}^\alpha(y)$ and $Y^A(\hat x)$ between $i^+\cup \scri^+$ and $i^-\cup \scri^-$, we obtain charges
\begin{equation}
 Q^{(\ln)}_\pm=  Q^{(\ln)}_{S,\pm}+ Q^{(\ln)}_{H,\pm}
\end{equation}
on the future (+) and past (-) boundary satisfying
\begin{equation}
    Q^{(\ln)}_+=Q^{(\ln)}_-.
\end{equation}
In the following we will establish that this classical conservation law, when recast as a symmetry of the S-matrix,
\begin{equation}\label{Qconservation}
    \bra{\text{out}}
    (Q^{(\ln)}_+ \mathcal S
    - \mathcal S Q^{(\ln)}_-)
    \ket{\text{in}} = 0,
\end{equation}
corresponds indeed to the classical logarithmic soft graviton theorem.
Before we proceed let us clarify some points. It may appear surprising that we have to turn to quantisation of the classical conservation law to show its equivalence with the {\it classical} logarithmic soft theorem. However, this is simply because the classical soft theorems are written in terms of asymptotic data of point particles as opposed to fields. Moreover, our classical charge derivation uses retarded propagators and we therefore recover the classical but not the quantum log soft theorem. To derive the latter the effect of quantum propagators would have to be taken into account and we leave this for future work.\footnote{Note that for the leading soft theorem (and all subleading tree-level soft theorems) the distinction between classical and quantum propagators do not matter.%
}

To prove that our superrotation charge conservation law implies the classical logarithmic soft graviton theorem we make use of the split into soft and hard charges. In turn we show that the commutator of the soft charge with the S-matrix gives rise to the soft graviton insertion, while the action of the hard charges on the matter fields lands us on the logarithmic soft graviton factor. Since the superrotation charges are smeared over the celestial sphere, to make contact with the soft theorem we make the following judicious choice for the superrotation vector field\footnote{This is the same vector field as was considered in \cite{Campiglia:2014yka} in deriving the subleading tree-level soft theorem from superrotation symmetry.} on $\cal I^+$,
\begin{equation}
    \label{gr_Ychoice}
    Y^z = \frac{(z-w)^2}{(\bar z-\bar w)},\qquad Y^\zb = 0
    .
\end{equation}
Using the identity $D_z^3Y^z = 4\pi \delta^{(2)}(z-w)$ this choice projects to negative-helicity gravitons; the log soft theorem derived via the Ward identity of this superrotation charge shall correspond to \eqref{Sln0Gclassical} with negative-helicity polarization tensor $\varepsilon^-_{\mu\nu}$.

\subsubsection*{Soft log charge insertion  $\braket{\text{out}|[Q^{(\ln)}_{S}, \mathcal S] |\text{in}}$}
The soft log charge for the choice of vector field \eqref{gr_Ychoice} takes the form
\begin{equation}
    Q_{S,+}^{(\ln)}[Y]
    =
        -\frac{8\pi}{\kappa^2}\gamma^{w\wb} \,     \mathcal J^{(\ln)}_{\wb\wb,+}
        + \frac{4\pi}{\kappa}\gamma^{w\wb}
        \overset0h_{rr}(w,\wb)
        {\cal J}^{(-1)}_{\wb\wb,+}.
\end{equation}
The sourced graviton, obtained from \eqref{hmn0T} via $\overset0h_{rr}=(\p_rx^\mu)(\p_rx^\nu)\overset0h_{\mu\nu}$, is given by
\begin{equation}
\badat2
    \overset0h_{rr}(w,\wb)
    &=
    -\frac{\kappa}{8\pi}\int d^3y \,(q(w,\wb)\cdot \Y(y)) \overset3T{}^{\rm matt}_{\tau\tau}(y)
    .
\eadat
\end{equation}
The matter energy-momentum tensor $\overset3T{}^{\rm matt}_{\tau\tau}$ is given by \eqref{T3tt}.
Upon quantization, the Poisson bracket \eqref{canonicalPB} becomes the following equal-time commutator for the ladder operators $b(y)$ and $b^\dagger(y)$,
\begin{equation}\label{bcommutator}
    [b(y),b^\dagger(y')]=(2\pi)^3\frac{2}{m^2}\delta^3(y-y')
    .
\end{equation}
Thus the action of $\overset0h_{rr}$ on an outgoing Fock state is simply
\begin{equation}
    \bra{\rm out}\overset0h_{rr}(w,\wb)
    =
        -\frac{\kappa}{8\pi}\sum_{i\in{\rm out}}(q(w,\wb)\cdot p_i)\bra{\rm out}
    .
\end{equation}
The insertion of the leading soft graviton operator ${\cal J}^{(-1)}_{\wb\wb,+}$ \cite{He:2014laa}
\begin{equation}
    {\cal J}^{(-1)}_{\wb\wb,+}
    =
        -\frac{\kappa}{4\pi(1+w\wb)^2}\lim_{\w\to0}
        \left[
            \w a_-(\w \hat x_w)
            + \w a_+^\dagger(\w\hat x_w)
        \right]
\end{equation}
yields the Weinberg soft graviton factor.
Here $a_\pm$ is the annihilation operator of a graviton with positive/negative helicity,
and $\hat x_w=\frac{1}{1+w\wb}(\wb+w,i(\wb-w),1-w\wb)$ is the unit 3-vector that points in the direction defined by $(w,\wb)$.
The insertion of the logarithmic soft graviton operator ${\cal J}^{(\ln)}_{\wb\wb,+}$ corresponds to a negative-helicty graviton insertion with the projector $\w\p_\w^2\w =\p_\w\w^2\p_\w$,
\begin{equation}
    {\cal J}^{(\ln)}_{\wb\wb,+}
    =
        \frac{i\kappa}{4\pi(1+w\wb)^2}
        \lim_{\w\to0}\p_\w \w^2 \p_\w\left[a_-(\w \hat x_w) - a_+^\dagger(\w \hat x_w)\right]
    .
\end{equation}
Following a similar analysis on the past boundary, we obtain the following expression for the soft charge insertion to the scattering amplitude,
\begin{equation}
\badat2
    \braket{\text{out}|(Q^{(\ln)}_{S,+} {\cal S - S} Q^{(\ln)}_{S,-})|\text{in}}
    &=
        -\frac{i}{\kappa}\lim_{\omega\to0}\p_\omega \omega^2 \p_\omega
        \bra{\text{out}}
            a_-^\text{out}(\omega\hat x_w){\cal S} - {\cal S}a_+^\text{in}(\omega \hat x_w)^\dagger
        \ket{\text{in}}
        \\&\quad
        + \frac{\kappa^2}{16\pi}\sum_{i\in{\rm out}} q\cdot p_i
        \sum_{j\in{\rm in,out}}\frac{(p_j\cdot \varepsilon^-)^2}{p_j\cdot q}
        \braket{\rm out|{\cal S}|\rm in}
    ,
    \label{gr_QsS-SQs}
\eadat
\end{equation}
where $\w q^\mu(w,\wb)$ is the momentum of the soft graviton, and in the second line we have used momentum conservation $\sum_{i\in{\rm in,out}} p_i=0$ to replace the sum over incoming momenta with the sum over outgoing ones.

\subsubsection*{Hard log charge insertion $\braket{\text{out}|[Q^{(\ln)}_{H}, \mathcal S] |\text{in}}$}
The hard logarithmic charge \eqref{grQHln} contains the `Coulombic' graviton operator $\overset1h_{\tau\tau}$.
It satisfies the differential equation \eqref{h1tt_diffeq}, so the inhomogeneous (sourced) solution can be written using the Green's function derived in Appendix~\ref{app:Green},
\begin{equation}
	\overset1h_{\tau\tau}(y)
	=
		\frac{\kappa}{8\pi}
        \int d^3y'  \frac{(\Y\cdot \Y')^2-\frac12}{\sqrt{(\Y\cdot \Y')^2-1}} \overset3T{}^{\rm matt}_{\tau\tau}(y')
	.\
\end{equation}
Here $\Y^\mu$ and $\Y'^\mu$ are the four-vectors parametrized by $y$ and $y'$ respectively as \eqref{Y}.
The matter energy-momentum tensor is given by \eqref{T3tt}.
Upon quantization, we obtain the following expression for the hard log charge, 
\begin{equation}\label{Qnormalordered}
    Q^{(\ln)}_H[\bar Y]
    =
        -\frac{\kappa^2 m^4}{4\pi}\Big(\frac{m}{4(2\pi)^3}\Big)^2
        {:}
		\int d^3y \,
        b^\dagger(y)b(y)
        \bar Y^\alpha
        \frac{\p}{\p y^\alpha}
        \int d^3y'
        \frac{(\Y\cdot \Y')^2-\frac12}{\sqrt{(\Y\cdot \Y')^2-1}} b^\dagger(y')b(y')
        {:}\,.
\end{equation}
The colons denote normal ordering with respect to the ladder operators $b_0$ and $b_0^\dagger$.
With the choice \eqref{gr_Ychoice} for the vector field, the map \eqref{YbarY} from $\cal I^+$ to $i^+$ implies that \cite{Campiglia:2015kxa}
\begin{equation}
    \bar Y^\alpha \frac{\p}{\p y^\alpha}
    =
        \frac{p\cdot \varepsilon^-}{p\cdot q}\varepsilon^-_\mu q_\nu
        \left(
        p^\mu \frac{\partial}{\partial p_\nu} - p^\nu \frac{\partial}{\partial p_\mu}
        \right)
    ,
\end{equation}
where $p^\mu = m\Y^\mu$ and $q^\mu=(1,\hat x_w)$.
Using this and the commutator \eqref{bcommutator}, we obtain the action of the future hard charge on an outgoing Fock state to be
\begin{align}
    \bra{\text{out}}
	Q^{(\ln)}_{H,+}
	&=
		-\frac{i\kappa^2}{16\pi}
		\sum_{\substack{i,j\in\text{out}\\i\neq j}}
		\frac{\varepsilon^-\cdot p_i}{p_i\cdot q}\varepsilon^-_\mu q_\nu
        \left[
        \left(
        p_i^\mu \frac{\partial}{\partial p_{i\nu}} - p_i^\nu \frac{\partial}{\partial p_{i\mu}}
        \right)
		\frac{(p_i\cdot p_j)^2-\frac12p_i^2p_j^2}{\sqrt{(p_i\cdot p_j)^2-p_i^2p_j^2}}
        \right]
		\bra{\text{out}}
    .
    \label{gr_QHinsertion_J}
\end{align}
The normal ordering of operators in \eqref{Qnormalordered} results in the sum in \eqref{gr_QHinsertion_J} being over \textit{distinct pairs} of outgoing scalars, in exact agreement with the soft factor \eqref{Sln0Gclassical} of Saha-Sahoo-Sen~\cite{Saha:2019tub}. %
Applying the momentum derivative, we land on the following expression
\begin{align}
    \bra{\text{out}}
	Q^{(\ln)}_{H,+}
	&=
		-\frac{\kappa^2}{32\pi}
		\sum_{\substack{i,j\in\text{out}\\i\neq j}}
		\frac{\varepsilon^-_{\mu\sigma} p_i^\sigma q_\nu}{p_i\cdot q}
		\frac{(p_i\cdot p_j)(p_i^\mu p_j^\nu-p_i^\nu p_j^\mu)}{[(p_i\cdot p_j)^2-p_i^2p_j^2]^{3/2}}
		[2(p_i\cdot p_j)^2-3p_i^2p_j^2]
		\bra{\text{out}}
	.
    \label{gr_QHinsertion}
\end{align}
When we repeat this analysis on the past time-like infinity, the action of the past hard charge $Q^{(\ln)}_{H,-}$ on the incoming state is analogous to \eqref{gr_QHinsertion} but with an overall minus sign.
Thus, the hard charge insertion amounts to
\begin{equation}\label{gr_QhS-SQh}
\badat2
    &
    \bra{\text{out}} (Q^{(\ln)}_{H,+} {\cal S - S} Q^{(\ln)}_{H,-} )\ket{\text{in}}
    \\&=
		-\frac{\kappa^2}{32\pi}
		\sum_{\substack{\eta_i\eta_j=1\\i\neq j}}
		\frac{\varepsilon^-_{\mu\sigma} p_i^\sigma q_\nu}{p_i\cdot q}
		\frac{(p_i\cdot p_j)(p_i^\mu p_j^\nu-p_i^\nu p_j^\mu)}{[(p_i\cdot p_j)^2-p_i^2p_j^2]^{3/2}}
		[2(p_i\cdot p_j)^2-3p_i^2p_j^2]
		\braket{\text{out}|{\cal S}|\text{in}}
    ,
\eadat
\end{equation}
where $\eta_i=+1$ ($-1$) if the $i$-th particle is outgoing (incoming).

\subsubsection*{Log charge conservation implies log soft theorem}

Collecting the two results \eqref{gr_QsS-SQs} and \eqref{gr_QhS-SQh} for the soft and hard charge insertions, we find that the classical conservation law \eqref{Qconservation} can be recast as the equation
\begin{equation}
\badat2
    &\lim_{\omega\to0}\p_\omega \omega^2 \p_\omega
    \bra{\text{out}}
        (a_-^\text{out}(\omega\hat x){\cal S} - {\cal S}a_+^\text{in}(\omega \hat x)^\dagger)
    \ket{\text{in}}
    \\ &=
        \frac{i(\frac\kappa2)^3}{4\pi}
		\sum_{\substack{\eta_i\eta_j=1\\i\neq j}}
		\frac{\varepsilon^-_{\mu\sigma} p_i^\sigma q_\nu}{p_i\cdot q}
		\frac{(p_i\cdot p_j)(p_i^\mu p_j^\nu-p_i^\nu p_j^\mu)}{[(p_i\cdot p_j)^2-p_i^2p_j^2]^{3/2}}
		[2(p_i\cdot p_j)^2-3p_i^2p_j^2]
		\braket{{\rm out}|{\cal S}|{\rm in}}
        \\&\quad
        - \frac{i(\frac\kappa2)^3}{2\pi}\sum_{i\in{\rm out}} q\cdot p_i
        \sum_{j\in{\rm in,out}}\frac{(p_j\cdot \varepsilon^-)^2}{p_j\cdot q}
        \braket{\rm out|{\cal S}|\rm in}
    .
\eadat
\end{equation}
The operator $\lim_{\w\to0}\p_\w \w^2\p_\w$ projects to the coefficient of $\ln\w$ in the soft expansion.
One recognizes the r.h.s.\ to be exactly the leading logarithmic soft factor \eqref{Sln0Gclassical} (times two, since the l.h.s.\ is a sum of two insertions).

We have thus derived, from first principles, a classical conservation law for superrotation symmetry that accounts for long-range infrared effects and established its equivalence with the classical logarithmic soft graviton theorem.
\acknowledgments

We would like to thank Samim Akhtar, Miguel Campiglia, Marc Geiller, Prahar Mitra, Ashoke Sen, Tom Wetzstein and Celine Zwikel for discussions. This work was supported by the Simons Collaboration on Celestial Holography. SC and AP are supported by the European Research Council (ERC) under the European Union’s Horizon 2020 research and innovation programme (grant agreement No 852386). 
This research was supported in part by Perimeter Institute for Theoretical Physics. Research at Perimeter Institute is supported by the Government of Canada through the Department of Innovation, Science and Economic Development and by the Province of Ontario through the Ministry of Research, Innovation and Science.

\newpage

\appendix

\section{Useful formulas}\label{app:UsefulFormulas}
\subsection*{Hyperbolic and Bondi coordinates}%

We list the metric components and the Christoffel symbols for the coordinates near the future time-like infinity $i^+$ and the future null infinity $\cal I^+$ that are used in the main text.

\paragraph{Time-like infinity.}
Near $i^+$, we employ the hyperbolic coordinates $(\tau,\rho,\hat x)$, where $\tau$ and $\rho$ are related to the Minkowski time $t$ and radial coordinate $r$ by
\begin{align}
    \tau = \sqrt{t^2 - r^2}
    ,\qquad
    \rho = \frac{r}{\sqrt{t^2-r^2}}
    ,\qquad
    t = \tau\sqrt{1+\rho^2}
    ,\qquad
    r=\tau\rho
    .
\end{align}
The line element in these coordinates takes the form
\begin{align}
    ds^2
    &=
        - d\tau^2
        + \tau^2 k_{\alpha\beta} dy^\alpha dy^\beta
    \label{appendix_i+_ds2}
    ,\\
    k_{\alpha\beta} dy^\alpha dy^\beta
    &=
       \frac{d\rho^2}{1+\rho^2}
       + \rho^2 \gamma_{AB} dx^A dx^B
    \label{appendix_k_ds2}
    ,
\end{align}
where $\gamma_{AB}$ is the unit sphere metric.
The metric $k_{\alpha\beta}$ describes a three-dimensional hyperboloid with unit negative curvature, and therefore the associated Ricci tensor is given by ${}^{(3)}{\cal R}_{\alpha\beta} = -2k_{\alpha\beta}$.
As a consequence, we have the following identities for any covariant vector $X_\alpha$ on the hyperboloid,
\begin{align}
    [\calD_\alpha,\calD^\beta] X_\beta
    &=
        2X_\alpha
    ,\\
    [\calD^\alpha,\calD^2] X_\alpha
    &=
        - 2\calD^\alpha X_\alpha
    ,
\end{align}
where $\calD_\alpha$ is the covariant derivative on the hyperboloid, compatible with $k_{\alpha\beta}$, and $\calD^2=\calD^\alpha\calD_\alpha$.
The only non-vanishing Christoffel symbols for the metric \eqref{appendix_i+_ds2} are
\begin{align}
    \Gamma^\tau_{\alpha\beta} = \tau k_{\alpha\beta}
    ,\qquad
    \Gamma^\alpha_{\tau\beta} = \frac{1}{\tau} \delta^\alpha_\beta
    ,\qquad
    \Gamma^\alpha_{\beta\gamma} = {}^{(3)}\Gamma^\alpha_{\beta\gamma}
    .
\end{align}
Here ${}^{(3)}\Gamma^\alpha_{\beta\gamma}$ denotes the Christoffel symbol for the metric \eqref{appendix_k_ds2} on the hyperboloid.
Written out explicitly, we have the following non-vanishing components:
\begin{equation}
\badat{2}
    \Gamma^\tau_{\rho\rho} = \frac\tau{1+\rho^2}
    ,\qquad
    \Gamma^\tau_{AB} = \rho^2 \tau \gamma_{AB}
    ,\qquad
    \Gamma^\rho_{\tau\rho} = \frac{1}{\tau}
    ,\qquad
    \Gamma^\rho_{\rho\rho} = -\frac\rho{1+\rho^2}
    ,\\
    \Gamma^\rho_{AB} = -\rho(1+\rho^2)\gamma_{AB}
    ,\qquad
    \Gamma^A_{\tau B} = \frac1\tau \delta^A_B
    ,\qquad
    \Gamma^A_{\rho B} = \frac1\rho \delta^A_B
    ,\qquad
    \Gamma^A_{BC} = {}^{(2)}\Gamma^A_{BC}
    ,
\eadat
\end{equation}
where ${}^{(2)}\Gamma^A_{BC}$ is the Christoffel symbol for the sphere metric $\gamma_{AB}$.\footnote{\label{sphereChristoffel}In the stereographic coordinates $\gamma_{z\zb}=\frac{2}{(1+z\zb)^2}$, we have ${}^{(2)}\Gamma^z_{zz} = \frac{-2\zb}{1+z\zb}$ and ${}^{(2)}\Gamma^\zb_{\zb\zb}=\frac{-2z}{1+z\zb}$.}
The Riemann tensor is given by the simple form
\begin{equation}
    {}^{(3)}\calR_{\alpha\beta\gamma\delta} = -k_{\alpha\gamma}k_{\beta\delta}
    + k_{\alpha\delta}k_{\beta\gamma}
    .
\end{equation}
~ %

\paragraph{Null infinity.}
Near $\cal I^+$, we employ the retarded time coordinate $(u,r,\hat x)$ where $u=t-r$.
In these coordinates, the Minkowski metric takes the form
\begin{align}
    ds^2
    &=
        -du^2 - 2dudr + r^2\gamma_{AB} dx^A dx^B
    ,
\end{align}
where $\gamma_{AB}$ is the unit sphere metric.
The only non-vanishing Christoffel symbols are
\begin{align}
    \Gamma^u_{AB} = r\gamma_{AB}
    ,\qquad
    \Gamma^r_{AB} = -r\gamma_{AB}
    ,\qquad
    \Gamma^A_{rB} = \frac1r \delta^A_B
    ,\qquad
    \Gamma^A_{BC} = {}^{(2)}\Gamma^A_{BC}
    ,
\end{align}
where ${}^{(2)}\Gamma^A_{BC}$ is the Christoffel symbol for $\gamma_{AB}$.\footnote{See footnote \ref{sphereChristoffel}.}

\subsection*{Equations of motion and gauge condition}
We summarize here the equations of motion for the real massive scalar field minimally coupled to gravity as well as the gauge conditions in, respectively, hyperbolic and retarded Bondi coordinates.

\paragraph{Time-like infinity.}
    The de Donder gauge condition \eqref{deDonder} in hyperbolic coordinates takes the form 
\begin{equation}
\label{deDondertaualpha}
\badat{2}
   0&=
		- \frac12\p_\tau h_{\tau\tau}
		+ \frac1{\tau^2}{\cal D}^\alpha h_{\alpha\tau}
		- \frac3\tau h_{\tau\tau}
		- \frac1{2\tau^2} k^{\alpha\beta}\p_\tau h_{\alpha\beta},\\
    0&=
        - \p_\tau h_{\tau\alpha}
		- \frac3\tau h_{\tau\alpha}
		+ \frac1{\tau^2}{\cal D}^\beta h_{\beta\alpha}
		- \frac12{\cal D}_\alpha h,
\eadat
\end{equation}
with the trace in hyperbolic coordinates given by $h=\frac1{\tau^2}k^{\alpha\beta}h_{\alpha\beta}
		-h_{\tau\tau}$.
The equation of motion \eqref{eom_gr_realscalar} for the real scalar field is given by
\begin{align}
\label{eomvarphi}
    0
	&=
		\Big[-\p_\tau^2 
		- \frac3\tau\p_\tau 
		+ \frac1{\tau^2}{\cal D}^2
		- m^2\Big]\varphi
		\nonumber\\&\quad
		+\kappa\Big[- h_{\tau\tau}
			\p_\tau^2 
		+ \frac{2}{\tau^2} h_{\tau\alpha}
		\Big(
			\p_\tau{\cal D}^\alpha 
			- \frac1\tau{\cal D}^\alpha
		\Big)
		- \frac{1}{\tau^4} h_{\alpha\beta}
		\left(
			{\cal D}^\alpha{\cal D}^\beta 
			- \tau k^{\alpha\beta} \p_\tau 
		\right)\Big]\varphi
	.
\end{align}
There is no radiation at $i^\pm$ but the matter stress tensor $ T^{\rm matt}_{\mu\nu}$ sources a `Coulombic' metric field via Einstein's equations which in hyperbolic coordinates take the form 
\begin{equation}
\badat{3}\label{Einsteinhyperbolic}
	-\frac{\kappa}{2}T_{\tau\tau}
	&=
		\frac12\p_\tau^2 \left(
			\frac1{\tau^2}k^{\alpha\beta}h_{\alpha\beta}
			+ h_{\tau\tau}
		\right)
		- \frac3{2\tau} \p_\tau h
		+ \frac1{2\tau^2}{\cal D}^2 h
		\\&\quad
		+ \frac3\tau \p_\tau h_{\tau\tau}
		+ \frac1{\tau^2}{\cal D}^2 h_{\tau\tau}
		- \frac2{\tau^2} \p_\tau {\cal D}^\alpha h_{\tau\alpha}
		- \frac4{\tau^4} k^{\alpha\beta}h_{\alpha\beta}
		+ \frac2{\tau^3} k^{\alpha\beta} \p_\tau h_{\alpha\beta}
    ,\\
	-\frac{\kappa}{2}T_{\tau\alpha}
	&=
		\p_\tau \left[
			\frac1{\tau^2} \left(
				k^{\beta\gamma} {\cal D}_\alpha h_{\beta\gamma}
				- {\cal D}^\beta h_{\alpha\beta}
			\right)
		\right]
		+ \frac2\tau {\cal D}_\alpha h_{\tau\tau}
		+ \frac1{\tau^2}({\cal D}^2-4) h_{\tau\alpha}
		- \frac1{\tau^2}{\cal D}^\beta{\cal D}_\alpha h_{\tau\beta}
    ,\\
	-\frac{\kappa}{2}T_{\alpha\beta}
	&=
		{\cal D}_\alpha{\cal D}_\beta h
		+ \frac1{\tau^2} \left(
			{\cal D}^2 h_{\alpha\beta}
			- {\cal D}^\gamma {\cal D}_\alpha h_{\gamma\beta}
			- {\cal D}^\gamma {\cal D}_\beta h_{\gamma\alpha}
			- 4 h_{\alpha\beta}
		\right)
		\\&\quad
		+ k_{\alpha\beta}\left[
			- \frac12 {\cal D}^2 h
			- 4 h_{\tau\tau}
			+ \frac12 \tau^2 \p_\tau^2 h
			+ \frac2\tau {\cal D}^\gamma h_{\tau\gamma}
			+ \tau\p_\tau\left(
				\frac12 h
				- 2 h_{\tau\tau}
			\right)
		\right]
		\\&\quad
		+ \frac1\tau \p_\tau h_{\alpha\beta}
		- \p_\tau^2 h_{\alpha\beta}
		+ \frac1\tau {\cal D}_\alpha h_{\tau\beta}
		+ \frac1\tau {\cal D}_\beta h_{\tau\alpha}
		+ \p_\tau \left(
			{\cal D}_\alpha h_{\tau\beta}
			+ {\cal D}_\beta h_{\tau\alpha}
		\right)
  .
\eadat
\end{equation}

\paragraph{Null infinity.}
In retarded Bondi coordinates the de Donder condition \eqref{deDonder} takes the form
\begin{equation}
\label{deDonderretarded}
\badat{3}
    0
    &=
        \left(\p_u + \p_r + \frac2r\right)h_{ur}
        - \left(\p_r + \frac2r\right)h_{uu}
        - \p_u h_{rr}
        + \frac{1}{r^2} D^A h_{uA}
        - \frac{1}{r^2}\gamma^{AB} \p_u h_{AB}
    ,\\
    0
    &=
        \left(\p_r - \frac2r\right) h_{ur}
        + \left(-\p_u + \frac2r\right) h_{rr}
        + \frac{1}{r^2} D^A h_{rA}
        + \frac{1}{r^2}\gamma^{AB}\left(-\p_r+\frac1r\right) h_{AB}
    ,\\
    0
    &=
        \left(\p_r - \p_u +\frac2r\right) h_{rA}
        - \left(\p_r +\frac2r\right) h_{uA}
        + D_A (2h_{ur}-h_{rr})
        + \frac{1}{r^2} \gamma^{BC}\left(
            D_B h_{AC}
            - D_A h_{BC}
        \right).
\eadat
\end{equation}
Einstein's equations \eqref{Einstein}, expressed in terms of the trace-reversed metric,
\begin{equation}
\label{hbar}
	\bar h_{\mu\nu} = h_{\mu\nu} - \frac12 \eta_{\mu\nu} h
	,
\end{equation}
are given by  $\Box \bar h_{\mu\nu} = -\frac{\kappa}{2}T_{\mu\nu}$ whose components take the form
\begin{equation}
\label{Einsteinretarded}
\badat6
    -\frac{\kappa}{2}T_{uu}
    &=
        \left[
            \p_r^2
            - 2\p_u\p_r
            + \frac2r(\p_r-\p_u)
            + \frac{1}{r^2}D^2
        \right]\bar h_{uu}
    ,\\
    -\frac{\kappa}{2}T_{ur}
    &=
        \left[
            \p_r^2
            - 2\p_u\p_r
            + \frac2r(\p_r-\p_u)
            + \frac{1}{r^2}D^2
        \right]\bar h_{ur}
        + \frac{2}{r^2}(\bar h_{uu}-\bar h_{ur})
        - \frac{2}{r^3}D^A\bar h_{uA}
    ,\\
    -\frac{\kappa}{2}T_{uA}
    &=
        \left[
            \p_r^2
            - 2\p_u\p_r
            + \frac{1}{r^2}(D^2-1)
        \right]\bar h_{uA}
        - \frac2r D_A(\bar h_{uu}-\bar h_{ur})
    ,\\
    -\frac{\kappa}{2}T_{rr}
    &=
        \left[
            \p_r^2
            - 2\p_u\p_r
            + \frac2r(\p_r-\p_u)
            + \frac{1}{r^2}D^2
        \right]\bar h_{rr}
        + \frac{4}{r^2}(\bar h_{ur}-\bar h_{rr})
        - \frac{4}{r^3}D^A\bar h_{rA}
        + \frac{2}{r^4}\gamma^{AB}\bar h_{AB}
    ,\\
    -\frac{\kappa}{2}T_{rA}
    &=
        \left[
            \p_r^2
            - 2\p_u\p_r
            + \frac{1}{r^2}(D^2-5)
        \right]\bar h_{rA}
        + \frac{4}{r^2}\bar h_{uA}
        - \frac2r D_A(\bar h_{ur}-\bar h_{rr})
        - \frac{2}{r^3}D^B\bar h_{AB}
    ,\\
    -\frac{\kappa}{2}T_{AB}
    &=
        \left[
            \p_r^2
            - 2\p_u\p_r
            - \frac2r(\p_r-\p_u)
            + \frac{1}{r^2}D^2
        \right]\bar h_{AB} + 2\gamma_{AB}(\bar h_{uu}+\bar h_{rr}-2\bar h_{ur})
        \\&\quad+ \frac2r \left[D_A(\bar h_{rB}-\bar h_{uB})+D_B(\bar h_{rA}-\bar h_{uA})\right].
\eadat
\end{equation}

\section{Tails from time-like to null infinity%
}
\label{app:tails}

The $u\to \infty$ behavior of the Cartesian components of the trace-reversed graviton field \eqref{hretardedfin} at $\scri^+$ is
\begin{align}\label{hbarmunu}
	 \bar h_{\mu\nu}(x)
	&=
		\frac1r\left[
			\overset0{\bar h}_{\mu\nu}(x^A)
			+ \frac{\ln u}{u} \overset{1,\ln}{\bar h_{\mu\nu}}(x^A)
			+ \frac1u \overset1{\bar h}_{\mu\nu}(x^A)
            + \cdots
		\right]
    ,
\end{align}
where the coefficients can be obtained using the expression \eqref{TmnCartesian} for the energy momentum tensor, 
\begin{align}
    \overset0{\bar h}_{\mu\nu}(x^A)
    &=
        \frac{\kappa}{8\pi}\int d^3y
            (-q\cdot\Y)^{-1}\Y_\mu \Y_\nu \overset3T{}_{\tau\tau}
    \nonumber,\\
    \overset{1,\ln} {{\bar h}_{\mu\nu}}(x^A)
    &=
        \frac{\kappa}{8\pi}\int d^3y
            \left(
                \Y_\mu \Y_\nu \overset{4,\ln}{T{}_{\tau\tau}}
                - \calD^\alpha(\Y_\mu \Y_\nu) \overset{3,\ln}{T_{\tau\alpha}}
            \right)
    \label{hlogu0}
    ,\\
    \overset1{\bar h}_{\mu\nu}(x^A)
    &=
        \frac{\kappa}{8\pi}\int d^3y
        \Bigg[
            \Y_\mu \Y_\nu \overset{4}T{}_{\tau\tau}
            - \calD^\alpha(\Y_\mu \Y_\nu) \overset{3}T_{\tau\alpha}
            + (\calD^\alpha\Y_\mu)(\calD^\beta\Y_\nu)\overset2T_{\alpha\beta}
            \\&\qquad\qquad
            -\ln(-q\cdot\Y)
            \left(
                \Y_\mu \Y_\nu \overset{4,\ln}{T{}_{\tau\tau}}
                - \calD^\alpha(\Y_\mu \Y_\nu) \overset{3,\ln}{T_{\tau\alpha}}
            \right)
        \Bigg]
    .
    \nonumber
\end{align}
The graviton energy-momentum tensor fall-offs \eqref{Thfalloff} imply that $\overset0h_{\mu\nu}$ and $\overset{1,\ln}{h_{\mu\nu}}$ receive contributions only from the matter energy-momentum tensor.
Now we use energy conservation $\nabla^\mu T_{\mu\tau}=0$, which in hyperbolic components translates to 
\begin{equation}
\label{EMconservation}
    -\left(\p_\tau+\frac3\tau\right) T_{\tau\tau}
    + \frac{1}{\tau^2} \calD^\alpha T_{\tau\alpha}
    - \frac{1}{\tau^3} k^{\alpha\beta} T_{\alpha\beta}
    =
        0.
\end{equation}
For the relevant terms in the $1/\tau$ expansion we have $\overset{4,\ln}{T_{\tau\tau}}+\mathcal D^\alpha\overset{3,\ln}{T_{\tau\alpha}}=0$ and $\overset{4}T_{\tau\tau}-\overset{4,\ln}{T_{\tau\tau}}+\mathcal D^\alpha\overset{3}T_{\tau\alpha}-k^{\alpha\beta}\overset2T_{\alpha\beta}=0$.\footnote{Note that while \eqref{EMconservation} together with the fall-offs \eqref{Tmattfalloff}, \eqref{Thfalloff} imply $\overset{4,\ln}{T_{\tau\tau}}+\mathcal D^\alpha\overset{3,\ln}{T_{\tau\alpha}}=0$ at $O(\kappa^3)$, we will show in Appendix \ref{app:Gravityallorder} that this actually holds to all orders.}
As a consequence the $\ln u/u$ term takes the form of an integral over a total derivative which vanishes,
\begin{equation}
     \overset{1,\ln}{{\bar h}_{\mu\nu}}(x^A)=-\frac{\kappa}{8\pi}\int d^3y\mathcal D^\alpha \left(\Y_\mu \Y_\nu \overset{3,\ln}{T_{\tau\alpha}}\right)=0.
\end{equation}
Now we use momentum conservation $\nabla^\mu T_{\mu\alpha}=0$, which in hyperbolic components translates to
\begin{equation}
    - \left(\p_\tau+\frac3\tau\right) T_{\tau\alpha}
    + \frac{1}{\tau^2} \calD^\beta T_{\alpha\beta}
    = 0
    .
\end{equation}
From the large-$\tau$ expansion we have $\overset{3,\ln}{T_{\tau\alpha}}-\calD^\beta\overset2T_{\alpha\beta}=0$.
The $1/u$ term simplifies to
\begin{equation}
\badat2
    \overset1{\bar h}_{\mu\nu}(x^A)
    &=
        \frac{\kappa}{8\pi}\int d^3y
        \left[
            - \Y_\mu \Y_\nu \calD^\alpha\overset{3,\ln}{T_{\tau\alpha}}
            + \ln(-q\cdot\Y)
                \calD^\alpha\left(\Y_\mu \Y_\nu \overset{3,\ln}{T_{\tau\alpha}}\right)
            + \left(
                \Y_\mu \Y_\nu k^{\alpha\beta}
                + (\calD^\alpha\Y_\mu)(\calD^\beta\Y_\nu)
            \right)\overset2T_{\alpha\beta}
        \right]
    \\ &=
        \frac{\kappa}{8\pi}\int d^3y\left[
        \frac{q^\sigma\Big(
                \Y_\sigma\calD^\alpha(\Y_\mu \Y_\nu)
                - \Y_\mu \Y_\nu \calD^\alpha \Y_\sigma
            \Big)}{(-q\cdot \Y)}
            \overset{3,\ln}{T_{\tau\alpha}}
            + \left(
                \Y_\mu \Y_\nu k^{\alpha\beta}
                + (\calD^\alpha\Y_\mu)(\calD^\beta\Y_\nu)
            \right)\overset2T_{\alpha\beta}
        \right]
    ,
\eadat
\end{equation}
where we have integrated by parts.
The graviton energy-momentum tensor \eqref{Thfalloff} does not contribute to $\overset{3,\ln}{T_{\tau\alpha}}$, and the matter energy-momentum tensor \eqref{Tmattfalloff} does not contribute to $\overset2T_{\alpha\beta}$.
Finally, we use $h_{\mu\nu}=\bar h_{\mu\nu}-\frac12\eta_{\mu\nu}\bar h$ and $\Y\cdot \Y=-1$ to obtain expressions for the coefficients in the large-$u$ expansion of $h_{\mu\nu}$ in \eqref{hmunu},
\begin{equation}
\badat4
    \overset0h_{\mu\nu}(x^A)
    &=
        \frac{\kappa}{8\pi}\int d^3y
            (-q\cdot\Y)^{-1}\left(\Y_\mu \Y_\nu+\frac12\eta_{\mu\nu}\right) \overset3T{}^{\rm matt}_{\tau\tau}
    ,\\
    \overset{1,\ln} {h_{\mu\nu}}(x^A)
    &=
        0
    ,\\
    \overset1h_{\mu\nu}(x^A)
    &=
        -\frac{\kappa}{8\pi}\int d^3y\bigg[
        \frac{
                (q\cdot\Y)\calD^\alpha(\Y_\mu \Y_\nu)
                - (\Y_\mu \Y_\nu+\frac12\eta_{\mu\nu}) \calD^\alpha (q\cdot\Y)
            }{q\cdot \Y}
            \overset{3,\ln}{T}{}^{\rm matt}_{\tau\alpha}
        \\&
            - \left(
                (\Y_\mu \Y_\nu+\frac12\eta_{\mu\nu}) k^{\alpha\beta}
                + (\calD^\alpha\Y_\mu)(\calD^\beta\Y_\nu)
                - \frac12\eta_{\mu\nu}(\calD^\alpha \Y_\sigma)(\calD^\beta\Y^\sigma)
            \right)\overset2T{}^h_{\alpha\beta}
        \bigg]
    .
\eadat
\end{equation}

\section{Green's functions and bulk-to-boundary propagators}\label{app:Green}
\subsection*{Vector Green's function $G^\alpha_A$}  %

The Green's function $G^\alpha_A(y;\hat x)$ used in \eqref{D2-2Green} is defined in terms of a scalar bulk-to-boundary propagator of the type $ {\cal D}^2 G^{(n)} = n(n-2) G^{(n)}$ with $n=4$, \cite{Campiglia:2015lxa}
\begin{align}
	G^\alpha_A(y;\hat x)\p_\alpha
	&=
		-G^{(4)}(y;\hat x)L_A^{\mu\nu}(\hat x)J_{\mu\nu}(y)
	,
\end{align}
with $L_A^{\mu\nu} = q^\mu D_A q^\nu-q^\nu D_Aq^\mu$ and $J_{\mu\nu} = x_\mu\p_\nu-x_\nu\p_\mu$. Here $q^\mu = (1,\hat x)$ and
\begin{equation}
    G^{(4)}(\rho,\hat x;\hat x')
    =
        \frac{3}{16\pi}\frac{1}{(-\Y\cdot q)^4},
\end{equation}
which satisfies
    $\lim_{\rho\to\infty} \rho^{-2}
    G^{(4)}(\rho,\hat x;\hat x') = \delta^{2}(\hat x-\hat x')$.

\subsection*{Green's function ${\cal G}$ for sourced graviton}

In this appendix we derive the Green's function satisfying the differential equation
\begin{align}\label{appD3eq}
	({\cal D}^2-3){\cal G}(y;y')
	&=
		-\delta^3(y-y')
	,
\end{align}
where $y$ and $y'$ are two points on a constant-$\tau$ hyperboloid.
Parametrizing the hyperboloid using the $\rho,\hat x$ coordinates in \eqref{rhox}, the delta function on the r.h.s.\ takes the form $\delta^3(y-y')=\frac{\sqrt{1+\rho^2}}{\rho^2}\delta(\rho-\rho')\delta^2(\hat x-\hat x')$.
In these coordinates the Laplacian takes the form
\begin{align}
	{\cal D}^2
	&=
		(1+\rho^2) \p_\rho^2
		+ \frac{(2+3\rho^2)}{\rho} \p_\rho
		+ \frac1{\rho^2}D^2
	,
\end{align}
where $D^2$ is the Laplacian on the unit sphere.
By symmetry, the solution to \eqref{appD3eq} only depends on the proper distance between two points on the hyperboloid. The proper distance squared is
\begin{align}
    (x-x')^2 = (\tau \Y - \tau \Y')^2 = 2\tau^2(-1-\Y\cdot \Y')
    ,
\end{align}
so we take ${\cal G}$ to be a single-variable function of $P\equiv -\Y\cdot \Y'$.
Note that $P\geq 1$, and $P=1$ corresponds to the contact point $y=y'$.
Without loss of generality, arrange the coordinates such that $y'$ sits at the origin.
This drops all angular dependence of $P$, and the map from $\rho$ to $P$ becomes simply $P=\sqrt{1+\rho^2}$.

Let us first solve the equation outside the source: $P>1$.
In this case the equation \eqref{appD3eq} becomes
\begin{align}
	(P^2-1){\cal G}''(P) + 3P{\cal G}'(P) - 3{\cal G}(P)
	&=
	0
	.
\end{align}
This has the following solution,
\begin{align}
	{\cal G}(P)
	&=
		A\frac{(P^2-\frac12)}{\sqrt{P^2-1}}
		- BP
    \label{Green0}
	,
\end{align}
where $A$ and $B$ are integration constants.
As this is a Green's function, it should be singular at the contact point $P=1$, so we set $B=0$.
We fix $A$ by going back to the differential equation.
Writing ${\cal G}$ in terms of $\rho$, we have
\begin{align}
	{\cal G}(y)
	&=
		A\frac{(1+2\rho^2)}{2\rho}
\end{align}
and the differential equation is
\begin{align}
	({\cal D}^2-3){\cal G}(y)
	&=
		-\frac{1}{4\pi}\frac{\sqrt{1+\rho^2}}{\rho^2}\delta(\rho)
	.
\end{align}
On the r.h.s.\ the angular delta function has been replaced by $\frac{1}{4\pi}$ since at $\rho=0$ both angles $\theta$, $\phi$ are degenerate.
The source is at the origin, so we can capture its contribution by integrating over any small volume around the origin and using Stokes' theorem.
Integrating both sides over an infinitesimal volume $\Sigma_\epsilon$ around the origin bounded by the surface $\rho=\epsilon$, we have
\begin{align}
	\lim_{\epsilon\to0}\int_{\Sigma_\epsilon} d^3y ({\cal D}^2-3){\cal G}(y)
	&=
		-1
	.
\end{align}
The integral of $-3{\cal G}$ can be done explicitly using the expression for ${\cal G}(y)$.
One finds that this contribution vanishes in the $\epsilon\to 0$ limit.
What remains is a total derivative, so we use Stokes' theorem to write
\begin{align}
	-1
	&=
		\lim_{\epsilon\to0}\int_{\Sigma_\epsilon} d^3y\,
		{\cal D}_\alpha {\cal D}^\alpha {\cal G}(y)
	=
		\left.
		\int d^2\hat x\,
		n^\alpha {\cal D}_\alpha {\cal G}(y)
		\right|_{\rho=0}
    .
\end{align}
The unit normal vector $n^\alpha$ has only one non-vanishing component $n^\rho=\rho^2\sqrt{1+\rho^2}$.
Plugging in the expression for ${\cal G}(y)$, we find that
\begin{align}
	-1
	&=
		\left.
		\int d^2\hat x\,
		n^\alpha {\cal D}_\alpha {\cal G}(y)
		\right|_{\rho=0}
	=
		-2\pi A
	,\qquad\implies\qquad
	A = \frac{1}{2\pi}
	.
\end{align}
Therefore, the Green's function for the differential operator ${\cal D}^2-3$ is
given by \eqref{Green0} with $A=\frac{1}{2\pi}$ and $B=0$, which with $P=-\Y\cdot \Y'$ reads
\begin{align}
	{\cal G}(y;y')
	&=
		\frac{1}{2\pi}\frac{(\Y\cdot \Y')^2-\frac12}{\sqrt{(\Y\cdot \Y')^2-1}}
	.
	\label{Green}
\end{align}

\section{%
Logarithmic charge to all orders in the coupling}
\label{app:Gravityallorder}

In this appendix, we show that the expressions \eqref{GravityQHln+} and  \eqref{GravityQSln+} for the logarithmic hard and soft charges in gravity are one-loop exact, in the sense that they do not receive further corrections at higher orders in $\kappa$.

To incorporate terms of higher power in the coupling constant, we allow for terms with higher powers of $\ln\tau$ in the large-$\tau$ expansion of the real massive scalar $\varphi$.
\begin{align}\label{app_phifalloff}
    \varphi(\tau,y)
    &=
        e^{-im\tau} \sum_{k=0}^\infty \sum_{n=0}^\infty \frac{(\ln\tau)^n}{\tau^{\frac32+k}} b_{k,n}(y) + \text{c.c.}~
    .
\end{align}
Matching with the notation used in the main text, we shall also refer to the first few coefficients as $b_k\equiv b_{k,0}$ and $\overset\ln b_k\equiv b_{k,1}$.
The graviton has falloffs given by \eqref{httfalloff} and \eqref{htahabfalloff} with one power of $\ln\tau$ in the next-to-leading order,
\begin{align}
\badat3\label{app_hfalloff}
    h_{\tau\tau}(\tau,y) &= \frac1\tau \overset1h_{\tau\tau}(y)+\frac{\ln\tau}{\tau^2}\overset\ln h_{\tau\tau}(y) + \frac{1}{\tau^2}\overset2h_{\tau\tau} \cdots
    ,\\
    h_{\tau\alpha}(\tau,y) &= \overset0h_{\tau\alpha}(y)+ \frac{\ln\tau}{\tau}\overset\ln h_{\tau\alpha}(y)+\cdots
    ,\\
    h_{\alpha\beta}(\tau,y) &= \tau \overset{-1}h_{\alpha\beta}(y)+\ln\tau\overset\ln h_{\alpha\beta}(y)+\cdots
    .
\eadat
\end{align}
It does not admit higher powers of log at this next-to-leading order in the large-$\tau$ expansion; such modes with $n$ factors of $\ln\tau$ are sourced by the $n$-th logarithmic modes $\overset{4,n}{T_{\tau\tau}}$, $\overset{3,n}{T_{\tau\alpha}}$ and $\overset{2,n}{T_{\alpha\beta}}$ of the energy-momentum tensor, but we shall see below that these vanish for all $n\geq 2$.

Since the leading falloffs of the graviton remain unchanged by higher order coupling corrections, the leading large-$\tau$ asymptotics of the graviton energy-momentum tensor \eqref{Th} also remain the same as the one-loop falloffs \eqref{Thfalloff},
\begin{equation}
    T^h_{\tau\tau} = \frac{1}{\tau^4}\overset4T{}^h_{\tau\tau} + \cdots
    ,\qquad
    T^h_{\tau\alpha} = \frac{1}{\tau^3}\overset3T{}^h_{\tau\alpha} + \cdots
    ,\qquad
    T^h_{\alpha\beta} = \frac{1}{\tau^2}\overset2T{}^h_{\alpha\beta} + \cdots
    ,
    \label{app_Thfalloff}
\end{equation}
to all orders in the coupling $\kappa$.

To find the large-$\tau$ behavior of the matter energy momentum tensor, we solve the scalar equation of motion to all orders in the coupling.
The scalar modes in \eqref{app_phifalloff} relevant for our purposes are ones up to order $\tau^{-\frac52}(\ln\tau)^n$, and for these it is sufficient to solve the equation \eqref{eom_gr_realscalar} with just one $\kappa^2$ correction\footnote{There are other corrections to the equation at this order in $\kappa$ coming from corrections to the covariant derivative, but they contribute to the equation only at order $\tau^{-\frac72}(\ln\tau)^n$ and higher in the large-$\tau$ limit.},
\begin{equation}
    \label{app_eomvarphi}
    \left[
        (g^{\mu\nu} - \kappa h^{\mu\nu} + \kappa^2 h^{\mu\sigma}h_\sigma^\nu+\cdots) \nabla_\mu\nabla_\nu - m^2
    \right]\varphi = 0
    ,
\end{equation}
which one may readily see by counting powers of $\tau$'s with the falloffs \eqref{app_phifalloff} and \eqref{app_hfalloff}.
Plugging the expansions \eqref{app_phifalloff} and \eqref{app_hfalloff} into \eqref{app_eomvarphi}, we find that the scalar equation of motion has the following large-$\tau$ expansion that is exact to all orders in $\kappa$:
\begin{align}
    0 &= e^{-im\tau}\sum_{n=0}^\infty \frac{(\ln\tau)^n}{\tau^{5/2}} {\rm Eq}_{\frac52,n}
    + e^{-im\tau}\sum_{n=0}^\infty \frac{(\ln\tau)^n}{\tau^{7/2}}{\rm Eq}_{\frac72,n}
    + \text{c.c.}
    + \cdots
    ,
\end{align}
where
\begin{align}
    {\rm Eq}_{\frac52,n}
    &=
        m^2\kappa \overset1h_{\tau\tau} b_{0,n}
        + 2im(n+1) b_{0,n+1}
    ,\\
    {\rm Eq}_{\frac72,n}
    &=
		\left(
			\calD^2
			+ \frac34
			- 3im\kappa \overset1h_{\tau\tau}
			- 2im\kappa \overset0h_{\tau\alpha}\calD^\alpha
			- im\kappa k^{\alpha\beta}\overset{-1}{h_{\alpha\beta}}
			+ m^2\kappa^2 (\overset1h_{\tau\tau})^2
			+ m^2\kappa \overset2h_{\tau\tau}
		\right)b_{0,n}
		\nonumber\\&\quad
		- 2im b_{1,n}
		+ 2im(n+1)b_{1,n+1}
		- (n+2)(n+1)b_{0,n+2}
		+ (n+1)b_{0,n+1}
		\nonumber\\&\quad
		+ m^2\kappa \overset1h_{\tau\tau}b_{1,n}
		+ 2im\kappa(n+1) \overset1h_{\tau\tau}b_{0,n+1}
		+ m^2\kappa\overset\ln h_{\tau\tau}b_{0,n-1}
    .
\end{align}
Solving ${\rm Eq}_{\frac52,n}= 0$ fixes all $b_{0,n}$ in terms of $b_0$,
\begin{equation}
    b_{0,n} = \frac{1}{n!} \left(\frac i2 m\kappa \overset1h_{\tau\tau}\right)^n b_0
    ,
    \label{grb0n}
\end{equation}
and $b_{0}$ is the unconstrained free data.
This implies that the leading ($k=0$) set of terms in \eqref{app_phifalloff} exponentiate to a phase,
\begin{equation}
    \varphi(\tau,y)
    =
        e^{-im\tau}\left[
        \frac{e^{\frac i2 m\kappa \overset1h_{\tau\tau}}}{\tau^{3/2}} b_0
        + \sum_{k=1}^\infty \sum_{n=0}^\infty \frac{(\ln\tau)^n}{\tau^{\frac32+k}} b_{k,n}(y) + \text{c.c.}
        \right]
    .
\end{equation}
Using \eqref{grb0n} to simplify expressions, we can write the equation ${\rm Eq}_{\frac72,n}=0$ as
\begin{align}
	b_{1,n}
	&=
		\frac{-i}{2m (1+iH)}\Bigg[
		\left(
			\calD^2
			+ \frac34
			- 5iH
			+ H^2
			- 2im\kappa \overset0h_{\tau\alpha}\calD^\alpha
			- im\kappa k^{\alpha\beta}\overset{-1}{h_{\alpha\beta}}
			+ m^2\kappa \overset2h_{\tau\tau}
		\right)\left(\frac{\left(iH\right)^n}{n!}b_0\right)
		\nonumber\\&\quad
		+ m^2\kappa\overset\ln h_{\tau\tau}b_{0,n-1}\Bigg]
		+ \frac{(n+1)}{1+iH}b_{1,n+1}
    ,
    \label{grb1neq}
\end{align}
where we define $b_{0,-1}=0$ for the special case $n=0$.
We have employed the shorthand
\begin{equation}\label{H}H\equiv \frac12 m\kappa \overset1h_{\tau\tau}\end{equation} to avoid notation clutter. The factor $(1+iH)^{-1}$ is a formal expression that represents the series
\begin{equation}
    \frac{1}{1+iH} \equiv \sum_{l=0}^\infty (-iH)^l
    .
\end{equation}
The equation \eqref{grb1neq} defines a recurrence relation, which allows us to write $b_{1,n}$ as
\begin{align}
    b_{1,n}
    &=
		\frac{-i}{2mn!}\sum_{k=n}^\infty \frac{1}{(1+iH)^{k-n+1}}
		\Bigg(
			\calD^2
			+ \frac34
			- 5iH
			+ H^2
			- 2im\kappa \overset0h_{\tau\alpha}\calD^\alpha
			- im\kappa k^{\alpha\beta}\overset{-1}{h_{\alpha\beta}}
			+ m^2\kappa \overset2h_{\tau\tau}
		\Bigg)\left(\left(iH\right)^kb_0\right)
		\nonumber\\&\quad
		+ \frac{m^2\kappa\overset\ln h_{\tau\tau} b_0}{2im n!}\sum_{k=n}^\infty \frac{k(iH)^{k-1}}{(1+iH)^{k-n+1}}    
    .
\end{align}
After applying the derivatives $\calD^2$ and $\calD^\alpha$, we find that each series is a Taylor expansion that reduces to a simple closed form:
\begin{align}
    b_{1,n}
    &=
		- \frac{1}{2imn!}[n(n-1)(iH)^{n-2} + 2n(iH)^{n-1}+2(iH)^n]
			(\calD^\alpha H)(\calD_\alpha H)b_0
		\nonumber\\&\quad
		+ \frac{1}{n!}[(iH)^n+n(iH)^{n-1}]\left(
			\frac{1}{m}(\p_\alpha H)\calD^\alpha
			+ \frac{1}{2m}(\calD^2 H) 
			- i\kappa\overset0h_{\tau\alpha}(\calD^\alpha H)
			- \frac i2 m\kappa\overset\ln h_{\tau\tau}
		\right)b_0
		\nonumber\\&\quad
		+ \frac{(iH)^n}{2imn!}
		\left(
			\calD^2
			+ \frac34
			- 5iH
                + H^2
			- im\kappa k^{\alpha\beta}\overset{-1}{h_{\alpha\beta}}
			+ m^2\kappa \overset2h_{\tau\tau}
			- 2im\kappa\overset0h_{\tau\alpha}\calD^\alpha
		\right)b_0
    .
\end{align}
Now we use this expression to derive the large-$\tau$ expansion of the relevant matter energy-momentum tensor components that is valid to all orders in the coupling constant.
The matter energy-momentum tensor is given by
\begin{equation}
    T^{\rm matt}_{\mu\nu}
    =
        \p_\mu \varphi \p_\nu \varphi
        - \frac12(g_{\mu\nu} + \kappa h_{\mu\nu})
        \left[
            (g^{\sigma\lambda} - \kappa h^{\sigma\lambda} + \kappa^2 h^{\sigma\kappa}h_\kappa^\lambda + O(h^3))\p_\sigma\varphi \p_\lambda\varphi + m^2\varphi^2
        \right]
    ,
\end{equation}
where $O(h^3)$ represents terms with three or more graviton factors.
Plugging the expansions \eqref{app_phifalloff} and \eqref{app_hfalloff}, we find that the energy-momentum tensor has the following large-$\tau$ expansions
\begin{equation}\label{app_Tmattfalloff0}
\badat2
    T^{\rm matt}_{\tau\tau}
    &=
        \sum_{n=0}^\infty \frac{(\ln\tau)^n}{\tau^3}\overset{3,n}T{}^{\rm matt}_{\tau\tau}
        + \sum_{n=0}^\infty \frac{(\ln\tau)^n}{\tau^4}\overset{4,n}T{}^{\rm matt}_{\tau\tau}
        + \cdots
    ,\\
    T^{\rm matt}_{\tau\alpha}
    &=
        \sum_{n=0}^\infty \frac{(\ln\tau)^n}{\tau^3}\overset{3,n}T{}^{\rm matt}_{\tau\alpha}
        + \cdots
    ,\\
    T^{\rm matt}_{\alpha\beta}
    &=
        \sum_{n=0}^\infty \frac{(\ln\tau)^n}{\tau^2}\overset{2,n}T{}^{\rm matt}_{\alpha\beta}
        + \cdots
    .
\eadat
\end{equation}
The coefficients take the form
\begin{align}
\begin{split}
    \overset{3,n}T{}^{\rm matt}_{\tau\tau}
    &=
        2m^2s^{(1)}_n
    ,\\
    \overset{2,n}T{}^{\rm matt}_{\alpha\beta}
    &=
        mk_{\alpha\beta}(2Hs^{(1)}_n + s^{(3)}_n)
    ,
\end{split}
\qquad
\begin{split}
    \overset{3,n}T{}^{\rm matt}_{\tau\alpha}
    &=
        ms^{(2)}_{n,\alpha}
    ,\\
    \overset{4,n}T{}^{\rm matt}_{\tau\tau}
    &=
        - 2mHs^{(1)}_n
        + ms^{(3)}_n
        + 2m^2s^{(4)}_n
    ,
\end{split}
\end{align}
given in terms of the four finite series
\begin{align}
\begin{split}
    s^{(1)}_n
    &=
        \sum_{k=0}^n b_{0,k}^* b_{0,n-k}
    ,\\
    s^{(3)}_n
    &=
        i\sum_{k=0}^{n+1}(n+1-2k)b_{0,k}^*b_{0,n+1-k}
\end{split}
\qquad
\begin{split}
    s^{(2)}_{n,\alpha}
    &=
        i\sum_{k=0}^n b_{0,n-k}^* \p_\alpha b_{0,k} + \text{c.c.},
    ,\\
    s^{(4)}_n
    &=
        \sum_{k=0}^n b_{1,k} b_{0,n-k}^*
        + \text{c.c.}
\end{split}
\end{align}
Let us evaluate these one by one:
\begin{itemize}
\item
    $s^{(1)}_n$ and $s^{(2)}_{n,\alpha}$:
    We find using the expression \eqref{grb0n} that both are simple binomial expansions of the form $(1-1)^{n-l}=\delta_{n,l}$, that reduce to either $\delta_{n,0}$ or $\delta_{n,1}$:
    \begin{equation}
    \badat2
        s^{(1)}_n
        &=
            |b_0|^2\delta_{n,0}
        ,\\
        s^{(2)}_n
        &=
            -2(\p_\alpha H)|b_0|^2\delta_{n,1}
            + i(b_0^*\p_\alpha b_0 - b_0\p_\alpha b_0^*)\delta_{n,0}
        .
    \eadat
    \end{equation}
\item
    $s^{(3)}_n$:
    With \eqref{grb0n}, the expression for $s_3$ can be written as a sum of two binomial expansions,
    \begin{equation}
    \badat2
        s^{(3)}_n
        &=
            i(iH)^{n+1}|b_0|^2\sum_{k=0}^{n+1}\frac{(n+1-2k)(-1)^k}{k!(n+1-k)!}
        \\ &=
        \frac{i}{n!}(iH)^{n+1}|b_0|^2
        \left[
            \sum_{k=0}^{n+1}\binom{n+1}{k}(-1)^k
            + 2\sum_{k=0}^{n}\binom{n}{k}(-1)^{k}
        \right]
        .
    \eadat
    \end{equation}
    The first term in square brackets is $\delta_{n+1,0}$ which is identically zero since $n$ is non-negative.
    The second term is simply $2\delta_{n,0}$.
    Thus,
    \begin{equation}
        s^{(3)}_n = -2H |b_0|^2 \delta_{n,0}
        .
    \end{equation}
\item
    $s^{(4)}_n$:
    This one involves both $b_{1,n}$ and $b_{0,n}$.
    Using \eqref{grb1neq} and \eqref{grb0n}, we find that each term that appear in the expression for $s^{(4)}_n$ is a simple binomial expansion that is proportional to either $\delta_{n,2}$, $\delta_{n,1}$ or $\delta_{n,0}$, and that the terms proportional to $\delta_{n,2}$ cancel out.
    We end up with
    \begin{align}
        s^{(4)}_n
        &=
            -\frac{\delta_{n,0}}{2m}
            \left[
                i b_0^*\calD^2 b_0
                + 5H|b_0|^2
                + m\kappa k^{\alpha\beta}\overset{-1}{h_{\alpha\beta}}|b_0|^2
                + 2m\kappa\overset0h_{\tau\alpha}b_0^*\calD^\alpha b_0
                + \text{c.c.}
            \right]
            \nonumber\\&\quad
            + \frac{1}{m}
            \left(
                \delta_{n,0}
                + \delta_{n,1}
            \right)
            \left[
                (\p_\alpha H)(b_0^*\calD^\alpha b_0 + b_0\calD^\alpha b_0^*)
                + (\calD^2 H)|b_0|^2 
            \right]
        .
    \end{align}
\end{itemize}
Putting the results together, we obtain
\begin{align}
    \overset{3,n}T{}^{\rm matt}_{\tau\tau}
    &=
        2m^2 b_0^* b_0 \delta_{n,0}
    ,\\
    \overset{3,n}T{}^{\rm matt}_{\tau\alpha}
    &=
        -m^2\kappa (\p_\alpha\overset1h_{\tau\tau}) b_0^* b_0 \delta_{n,1}
        + im(b_0^*\p_\alpha b_0 - b_0 \p_\alpha b_0^*)\delta_{n,0}
    ,\\
    \overset{2,n}T{}^{\rm matt}_{\alpha\beta}
    &=
        0
    ,
\end{align}
and
\begin{align}\label{gr_4nTtt}
    \overset{4,n}T{}^{\rm matt}_{\tau\tau}
    &=
        -m\delta_{n,0}
        \left[
            i(b_0^*\calD^2b_0 - b_0\calD^2 b_0^*)
            + m\kappa(
                7\overset1h_{\tau\tau}
                + 2k^{\alpha\beta}\overset{-1}{h_{\alpha\beta}}
            )|b_0|^2
            + 2m\kappa\overset0h_{\tau\alpha}(b_0^*\calD^\alpha b_0+b_0\calD^\alpha b_0^*)
        \right]
        \nonumber\\&\quad
        + m^2\kappa\left(
            \delta_{n,0}
            + \delta_{n,1}
        \right)
        \left[
            (\p_\alpha \overset1h_{\tau\tau})(b_0^*\calD^\alpha b_0 + b_0\calD^\alpha b_0^*)
            + (\calD^2 \overset1h_{\tau\tau})|b_0|^2 
        \right]
    .
\end{align}
In particular, combining these with the falloffs \eqref{app_Thfalloff} of the graviton energy-momentum tensor, we find that the total energy-momentum tensor satisfies
\begin{equation}
    0
    =
        \overset{4,n}T{}_{\tau\tau}
    =
        \overset{3,n}T{}_{\tau\alpha}
    =
        \overset{2,n}T{}_{\alpha\beta}
    \qquad\text{for all $n\geq2$}
    .
\end{equation}
These are the modes that source the graviton modes $\overset{2,n\geq2}{h_{\tau\tau}}$, $\overset{1,n\geq2}{h_{\tau\alpha}}$, $\overset{0,n\geq2}{h_{\alpha\beta}}$ with two or more powers of $\ln\tau$ at the next-to-leading order in $1/\tau$; the fact that they vanish justifies the graviton falloff condition \eqref{app_hfalloff}.

Therefore, to all orders in the coupling constant, the large-$\tau$ behavior of the energy-momentum tensor components $T_{\tau\tau}$ and $T_{\tau\alpha}$ is given by
\begin{equation}\label{app_Tmattfalloff}
\badat3
    T^{\rm matt}_{\tau\tau}
    &=
        \frac{1}{\tau^3}\overset{3}T{}^{\rm matt}_{\tau\tau}
        + \frac{\ln\tau}{\tau^4}\overset{4,\ln}T{}^{\rm matt}_{\tau\tau}
        + \frac{1}{\tau^4}\overset4T{}^{\rm matt}_{\tau\tau}
        + \cdots
    ,\\
    T^{\rm matt}_{\tau\alpha}
    &=
        \frac{\ln\tau}{\tau^3}\overset{3,\ln}T{}^{\rm matt}_{\tau\alpha}
        + \frac{1}{\tau^3}\overset3T{}^{\rm matt}_{\tau\alpha}
        + \cdots
    ,\\
    T^{\rm matt}_{\alpha\beta}
    &=
        \frac{1}{\tau^2}\overset2T{}^{\rm matt}_{\alpha\beta}
        + \cdots
\eadat
\end{equation}
where
\begin{align}
    \overset3T{}^{\rm matt}_{\tau\tau}
    &=
        2m^2 b_0^* b_0 \delta_{n,0}
    \label{gr_30Ttt}
    ,\\
    \overset{3,\ln}T{}^{\rm matt}_{\tau\alpha}
    &=
        -m^2\kappa (\p_\alpha\overset1h_{\tau\tau}) b_0^* b_0
    \label{gr_31Tta}
    ,\\
    \overset3T{}^{\rm matt}_{\tau\alpha}
    &=
        im(b_0^*\p_\alpha b_0 - b_0 \p_\alpha b_0^*)
    \label{gr_30Tta}
    ,\\
    \overset2T{}^{\rm matt}_{\alpha\beta}
    &=
        0
    .
\end{align}
The subleading coefficients $\overset{4,\ln}T{}^{\rm matt}_{\tau\tau}$, $\overset4T{}^{\rm matt}_{\tau\tau}$ are given by the coefficients of $\delta_{n,1}$ and $\delta_{n,0}$ in \eqref{gr_4nTtt} respectively.
What is important for us is the fact that $T{}^{\rm matt}_{\tau\tau}$ does not develop higher powers of logs of the form $\tau^{-4}(\ln\tau)^n$ with $n\geq 2$, to all orders in the coupling; their explicit form is not of interest in this paper.

\subsection*{Logarithmic hard charge %
}

The logarithmic hard charge is given by
\begin{equation}
    Q^{(\ln)}_{H,+}[\bar Y] = \int_{i^+}d^3y\,\bar Y^\alpha\overset{3,\ln} {T_{\tau\alpha}}
    .
\end{equation}
The energy-momentum tensor has matter and graviton contributions,
\begin{equation}
    T_{\tau\alpha} = T^{\rm matt}_{\tau\alpha} + T^h_{\tau\alpha}
    .
\end{equation}
The graviton part \eqref{Thfalloff} is of order $O(\kappa^2)$, and goes as $\tau^{-3}$,
\begin{equation}
    T^h_{\tau\alpha}
    =
        \frac{1}{\tau^3}\overset3T{}^h_{\tau\alpha} + \cdots
    ,
\end{equation}
so it does not contribute to the log hard charge.
The matter part takes the form \eqref{gr_31Tta}, which implies that the log hard charge is given by
\begin{equation}
    Q_{H,+}^{(\ln)}[\bar Y]
    =
        - m^2\kappa \int_{i^+} d^3y\,
        \bar Y^\alpha(\p_\alpha\overset1h_{\tau\tau}) b_0^*b_0.
\end{equation}
This expression, which is exact to all orders in $\kappa$, is in agreement with \eqref{grQHln} and \eqref{T3lntamatt}, which demonstrates that the hard log charge of gravity is one-loop exact.

\subsection*{Subleading `tree-level' hard charge %
}

The `tree-level' hard charge takes the form
\begin{equation}
    Q^{(0)}_{H,+}[\bar Y] = \int_{i^+}d^3y\,\bar Y^\alpha\overset{3} T_{\tau\alpha}
    .
\end{equation}
The coefficient $\overset3T_{\tau\alpha}$ receives contributions from both matter and gravitons,
\begin{equation}
    \overset3T_{\tau\alpha} = \overset3T{}^{\rm matt}_{\tau\alpha} + \overset3T{}^h_{\tau\alpha}
    .
\end{equation}
The graviton contribution is of order $O(\kappa^2)$, while the matter term \eqref{gr_30Tta} is of order $\kappa^0$; it does not get corrections at higher order in $\kappa$. Thus
\begin{equation}
    Q^{(0)}_{H,+}[\bar Y] = im\int_{i^+}d^3y\,\bar Y^\alpha(b_0^*\p_\alpha b_0 - b_0 \p_\alpha b_0^*)
    + O(\kappa^2)
    .
\end{equation}
The higher-order correction comes only from the graviton energy-momentum tensor.

\subsection*{Logarithmic soft charge %
}

Higher powers of $\ln\tau$ in the energy-momentum imply that we should consider a generalization of the large-$u$ expansion of the graviton field \eqref{hmunu},
\begin{align}\label{hmunu_generalization}
	h_{\mu\nu}(x)
	&=
		\frac1r\left[
			\overset0h_{\mu\nu}(x^A)
			+ \sum_{n=1}^\infty \frac{(\ln u)^n}{u} \overset{1,n}{h_{\mu\nu}}(x^A)
			+ \frac1u \overset1h_{\mu\nu}(x^A)
            + \cdots
		\right]
    .
\end{align}
The coefficients in the expansion are related to those appearing in the generalization of \eqref{hbarmunu} for the trace-reversed field by $h_{\mu\nu}=\bar h_{\mu\nu}-\frac12\eta_{\mu\nu}\bar h$,
\begin{align}\label{hbarmunu_generalization}
	{\bar h}_{\mu\nu}(x)
	&=
		\frac1r\left[
			\overset0{\bar h}_{\mu\nu}(x^A)
			+ \sum_{n=1}^\infty \frac{(\ln u)^n}{u} \overset{1,n}{{\bar h}_{\mu\nu}}(x^A)
			+ \frac1u \overset1{\bar h}_{\mu\nu}(x^A)
            + \cdots
		\right]
    ,
\end{align}
where
\begin{align}
	\overset0{\bar h}_{\mu\nu}
	&=
		\frac{\kappa}{8\pi}\int d^3y
            (-q\cdot\Y)^{-1}
            \Y_\mu \Y_\nu \overset3T{}_{\tau\tau}
    \nonumber,\\
	\overset{1,n}{{\bar h}_{\mu\nu}}
	&=
        \frac{\kappa}{8\pi}\int d^3y
        \sum_{r=n}^\infty \binom{r}{n}[-\ln(-q\cdot\Y)]^{r-n}
        \nonumber\\&\qquad\qquad\times
        \left(
            \Y_\mu \Y_\nu \overset{4,r}{T{}_{\tau\tau}}
            - \calD^\alpha(\Y_\mu \Y_\nu) \overset{3,r}{T_{\tau\alpha}}
            + (\calD^\alpha\Y_\mu)(\calD^\beta\Y_\nu)\overset{2,r}{T_{\alpha\beta}}
        \right)
        ,
\end{align}
and $\overset1{\bar h}_{\mu\nu} \equiv \overset{1,0}{{\bar h}_{\mu\nu}}$.
Since we saw in \eqref{app_Tmattfalloff} and \eqref{app_hfalloff} that both $\overset{4,r}{T_{\tau\tau}}$ and $\overset{3,r}{T_{\tau\alpha}}$ vanish for all $r\geq 2$ and that $\overset{2,r}{T_{\alpha\beta}}$ vanish for all $r$ except for $\overset{2,0}{T_{\alpha\beta}}=\overset2T{}^h_{\alpha\beta}$, we have
\begin{align}
    \overset{1,n}{{\bar h}_{\mu\nu}} &= 0 \qquad\text{for all $n\geq 2$}
    \nonumber,\\
	\overset{1,1}{{\bar h}_{\mu\nu}}
	&=
		\frac{\kappa}{8\pi}\int d^3y
			\left(
				\Y_\mu \Y_\nu \overset{4,1}{T{}_{\tau\tau}}
				- \calD^\alpha(\Y_\mu \Y_\nu) \overset{3,1}{T_{\tau\alpha}}
			\right)
    ,\\
    \overset1{\bar h}_{\mu\nu}
	&=
		\frac{\kappa}{8\pi}\int d^3y
        \Bigg[
            \Y_\mu \Y_\nu \overset{4}{T{}_{\tau\tau}}
            - \calD^\alpha(\Y_\mu \Y_\nu) \overset{3}{T_{\tau\alpha}}
            + (\calD^\alpha\Y_\mu)(\calD^\beta\Y_\nu)\overset2T_{\alpha\beta}
            \nonumber\\&\qquad\qquad
            -\ln(-q\cdot\Y)
            \left(
                \Y_\mu \Y_\nu \overset{4,1}{T{}_{\tau\tau}}
                - \calD^\alpha(\Y_\mu \Y_\nu) \overset{3,1}{T_{\tau\alpha}}
            \right)
        \Bigg].
    \nonumber
\end{align}
For $\overset{1,1}{{\bar h}_{\mu\nu}}$, we can integrate by parts and use the energy-momentum conservation $\nabla^\nu T_{\tau\nu} = 0$ which reads
\begin{equation}
    \overset{4,n}{T_{\tau\tau}} + \calD^\alpha \overset{3,n}{T_{\tau\alpha}} = (n+1)\overset{4,n+1}{T_{\tau\tau}}
    + k^{\alpha\beta}\overset{2,n}{T_{\alpha\beta}}
\end{equation}
to write the integrand as $\overset{4,2}T_{\tau\tau}$ which is zero.
Therefore the $u^{-1}\ln u$ mode vanishes as well:
\begin{align}
	\overset{1,1} {\bar h}_{\mu\nu}
	&=
		\frac{\kappa}{8\pi}\int d^3y
			\Y_\mu \Y_\nu 
			\left(
				\overset{4,1}{T{}_{\tau\tau}}
				+\calD^\alpha \overset{3,1}{T_{\tau\alpha}}
			\right)
	=
		\frac{\kappa}{8\pi}\int d^3y
			\Y_\mu \Y_\nu 
			\left(2\overset{4,2}{T{}_{\tau\tau}}+k^{\alpha\beta}\overset{2,1}{T_{\alpha\beta}}\right)
	= 0
    .
\end{align}
This establishes $\overset{1,n}{{\bar h}_{\mu\nu}}=0$ for $n\geq1$, and therefore $\overset{1,n}{h_{\mu\nu}}=0$ for all $n\geq1$, to all orders in the coupling: the expansion \eqref{hmunu_generalization} does not admit corrections of the form $\frac{(\ln u)^n}{u}$ with $n\geq 1$ even at higher powers in the coupling.

\bibliographystyle{jhep}
\bibliography{references}

\end{document}